\documentclass[fleqn,usenatbib,useAMS]{mnras}

\usepackage[T1]{fontenc}
\usepackage{graphicx}
\DeclareGraphicsExtensions{.pdf,.png}
\usepackage{color}
\usepackage{amsmath}
\usepackage{amssymb}
\usepackage{comment}
\usepackage{amsfonts}
\usepackage[normalem]{ulem}
\usepackage{footnote}
\usepackage{widetext}
\usepackage{dsfont}
\usepackage{enumitem}

\DeclareFontFamily{OT1}{pzc}{}
\DeclareFontShape{OT1}{pzc}{m}{it}{<-> s * [1.10] pzcmi7t}{}
\DeclareMathAlphabet{\mathpzc}{OT1}{pzc}{m}{it}

\usepackage{fancyvrb}
\usepackage{fancybox}

\definecolor{blue}{rgb}{0,0,1}
\definecolor{darkergreen}{rgb}{0,0.5,0}

\defcitealias{Planck2018PNG}{Planck Collaboration (2019)}

\newcommand{\eg}{e.g$.$ }
\newcommand{\Eg}{E.g$.$ }

\newcommand{\ie}{i.e$.$ }
\newcommand{\cf}{cf$.$ }

\newcommand{\wrt}{wrt$.$ }
\newcommand{\dd}{\ensuremath{\mathrm{d}}}

\newcommand{\tabref}[1]{Table \ref{tab:#1}}

\newcommand{\figref}[1]{Figure \ref{fi:#1}}
\newcommand{\figrefnospace}[1]{Figure \ref{fi:#1}}
\newcommand{\Figref}[1]{Figure \ref{fi:#1}}

\newcommand{\eqnref}[1]{Equation \ref{eq:#1}}
\newcommand{\eqnrefnospace}[1]{Equation \ref{eq:#1}}

\newcommand{\appref}[1]{Appendix \ref{app:#1}}
\newcommand{\apprefnospace}[1]{Appendix \ref{app:#1}}

\newcommand{\secref}[1]{Section \ref{sec:#1}}
\newcommand{\secrefnospace}[1]{Section \ref{sec:#1}}
\newcommand{\Secref}[1]{Section \ref{sec:#1}}

\allowdisplaybreaks

\title[Measuring $f_{\mathrm{NL}}$ with the bulk of the density PDF]{Primordial non-Gaussianity without tails - how to measure $f_{\mathrm{NL}}$ with the bulk of the density PDF}
\author[Oliver Friedrich et al.]{\parbox{\linewidth}{Oliver Friedrich$^{1, 2}$, Cora Uhlemann$^{3,4}$, Francisco Villaescusa-Navarro$^{5,6}$, Tobias Baldauf$^{3,7}$, Marc Manera$^{3,8}$, Takahiro Nishimichi$^{9,10}$}
\vspace*{10pt}\\
$^1$ {Kavli Institute for Cosmology, University of Cambridge, CB3 0HA Cambridge, United Kingdom}\\
$^{2}$ {Churchill College, University of Cambridge, CB3 0DS Cambridge, United Kingdom}\\
$^{3}$ {Centre for Theoretical Cosmology, DAMTP, University of Cambridge, CB3 0WA Cambridge, United Kingdom}\\
$^{4}$ {Fitzwilliam College, University of Cambridge, CB3 0DG Cambridge, United Kingdom}\\
$^5$ Department of Astrophysical Sciences, Princeton University, Peyton Hall, Princeton NJ 08544-0010, USA\\
$^6$ Center for Computational Astrophysics, Flatiron Institute, 162 5th Avenue, 10010, New York, NY, USA\\
$^7$ {Clare Hall, University of Cambridge, CB3 9AL Cambridge, United Kingdom}\\
$^8$ Institut de F\'isica d'Altes Energies (IFAE), Barcelona Institute of Science and Technology, Campus UAB, 08193 Bellaterra (Barcelona) Spain\\
$^9$ Center for Gravitational Physics, Yukawa Institute for Theoretical Physics, Kyoto University, Kyoto 606-8502, Japan\\
$^{10}$ Kavli Institute for the Physics and Mathematics of the Universe (WPI), University of Tokyo, Kashiwa, Chiba 277-8583, Japan\\
}
\date{December 2019}

\begin{document}

\thisfancyput(14.8cm,0.5cm){\large{YITP-19-122}}

\maketitle

\begin{abstract}
We investigate the possibility to detect primordial non-Gaussianity by analysing the bulk of the probability distribution function (PDF) of late-time cosmic density fluctuations. For this purpose we devise a new method to predict the impact of general non-Gaussian initial conditions on the late-time density PDF. At redshift $z=1$ and for a smoothing scale of $30$Mpc/$h$ our predictions agree with the high-resolution Quijote N-body simulations to $\sim 0.2\%$ precision. This is within cosmic variance of a $\sim 100(\mathrm{Gpc}/h)^3$ survey volume. When restricting to this $30$Mpc/$h$ smoothing scale and to mildly non-linear densities ($\delta[30\mathrm{Mpc}/h] \in [-0.3, 0.4]$) and also marginalizing over potential ignorance of the amplitude of the non-linear power spectrum an analysis of the PDF for such a survey volume can still measure the amplitude of different primordial bispectrum shapes to an accuracy of \smash{$\Delta f_{\mathrm{NL}}^{\mathrm{loc}} = \pm 7.4\ ,\ \Delta f_{\mathrm{NL}}^{\mathrm{equi}} = \pm 22.0\ ,\ \Delta f_{\mathrm{NL}}^{\mathrm{ortho}} = \pm 46.0\ $}. When pushing to smaller scales and assuming a joint analysis of the PDF with smoothing radii of $30$Mpc/$h$ and $15$Mpc/$h$ ($\delta[15\mathrm{Mpc}/h] \in [-0.4, 0.5]$) this improves to \smash{$\Delta f_{\mathrm{NL}}^{\mathrm{loc}} = \pm 3.3\ ,\ \Delta f_{\mathrm{NL}}^{\mathrm{equi}} = \pm 11.0\ ,\ \Delta f_{\mathrm{NL}}^{\mathrm{ortho}} = \pm 17.0\ $} - even when marginalizing over the non-linear variances at both scales as two free parameters. Especially, such an analysis could simultaneously measure $f_{\mathrm{NL}}$ and the amplitude and slope of the non-linear power spectrum. However, at $15$Mpc/$h$ our predictions are only accurate to $\lesssim 0.8\%$ for the considered density range. We discuss how this has to be improved in order to push to these small scales and make full use of upcoming surveys with a PDF-based analysis.
\end{abstract}


\section{Introduction}

Data of the large scale structure of the Universe can be successfully analysed on the basis of the 1-point probability distribution function (PDF) of the matter density field - even in the presence of tracer bias and redshift uncertainties. This has been demonstrated \eg by \citet{Gruen2016, Friedrich2018, Gruen2018, Brouwer2018}. In particular, \citet{Gruen2018} and \citet{Friedrich2018} measured the PDF of galaxy density and then used measurements of gravitational lensing to relate that to the PDF of the underlying matter density field quantile-by-quantile. This way they could simultaneously
\begin{itemize}
    \item[a)] test the $\Lambda$CDM prediction for how the variance and skewness of matter density fluctuations are related on mildly non-linear scales
    \item[b)] constrain a two parameter galaxy bias model, that accounts for both linear bias and density dependent shot-noise
    \item[c)] measure the late-time matter density and the amplitude of late-time density fluctuations as encoded by the parameters $\Omega_m$ and $\sigma_8$ of the $\Lambda$CDM model.
\end{itemize}

Given the rich amount of information that can be harvested from the PDF \citep[see also][]{Uhlemann2020}, it is time to explore its potential for constraining fundamental physics and to compare it to other cosmological probes. In this paper we showcase one specific application: we study how primordial non-Gaussianity \citep[see \eg][and references therein]{Komatsu2001, Fergusson2009, Scoccimarro2012, Meerburg2019a, Biagetti2019} are imprinted in the late time density PDF and how constraints from such an analysis compare to the ones obtained from direct measurements of moments of the density field. The impact of primordial non-Gaussianity on the matter density PDF has previously been discussed \eg by \citet{Valageas2002III, Uhlemann2018b}. We extend on their results in two ways: First, we present a new method to model the impact of general non-Gaussian initial conditions on the PDF of the late-time density field. This method directly models the cumulant generating function (CGF) of the late-time density field from the CGF of the early-time density field. As shown in \secref{model} such an approach requires fewer approximating steps than existing modelling approaches and is close to what would be called modelling 'from first principles'. Secondly, we take into account the full covariance matrix of measured density PDFs across different density contrasts and for two different smoothing scales to determine how well measurements of the density PDF can determine the amplitude of different primordial bispectrum templates. In the context of this task, we also compare the statistical power of the density PDF to that of direct measurements of the cumulants of the density field. The latter have recently been pushed towards applicability in real large scale structure analyses by \citet{Gatti2019} and the impact of primordial non-Gaussianity on higher-order weak lensing statistics has \eg been investigated by \citet{Pace2011}.

In general, scale-dependent tracer bias is believed to be the most promising signature of local primordial non-Gaussianity in the large-scale structure \citep[\eg][]{Dalal2008,Desjacques2009, Jeong2009, Scoccimarro2012,Biagetti2017}, especially when combined with cosmic variance cancellation techniques \citep{Seljak2009}. Recently, it was pointed out that a similar scale-dependent bias effect from primordial non-Gaussianity can be observed with voids \citep{Chan19}, although massive neutrinos produce a similar effect on scales smaller than their maximal comoving free streaming scale \citep{Banerjee2016}. Detecting primordial non-Gaussianity in scale-dependent tracer bias requires analyses of clustering power spectra at very large scales. This poses a challenge in terms of cosmic variance, systematic effects \citep{Laurent2017} as well as modelling of large scale relativistic effects \citep{Bartolo2011, Camera2015, Contreras2019}. A way to measure primordial non-Gaussianity that does not suffer from these challenges (but instead from other ones) is to probe the PDF of densities in spheres and their density-dependent clustering on intermediate scales \citep{Uhlemann2018b, Codis2016b}. In fact, the density PDF is sensitive to all primordial bispectrum shapes and can hence probe equilateral or orthogonal templates for which scale-dependent bias is less pronounced.
In addition, density-dependent clustering allows to disentangle local $f_{\rm NL}$ (causing primordial skewness) and $g_{\rm NL}$ (generating primordial kurtosis) by scanning different density environments. Studying the 1-point PDF parallels a number of efforts to understand the cosmic structures beyond their $N$-point statistics - both for the purpose of detecting primordial non-Gaussianity (\eg \citealt{Chiang2015}; \citealt{Nusser2018}; \citealt{Karagiannis2019}; \citealt{MoradinezhadDizgah2019}) and to test the theory of structure formation in general (\eg \citealt{Jain2000}; \citealt{Simpson2013}; \citealt{Codis2016a, Kacprzak2016}; \citealt{Coulton2019}).

Eventhough there are numerous ways how primordial non-Gaussianity can emerge from inflation, one can categorise them according to the primordial bispectrum shape they generate \citep[see \eg][for a discussion of concrete models]{Babich2004,Chen2007,Liguori2010}. As suggested by its name, the local shape is typically generated by local interactions, such as in multi-field inflation \citep{BernardeauUzan2002} or curvaton models \citep{Bartolo2004}, with a small amplitude also being produced in single-field slow-roll inflation \citep{Acquaviva2003}. The equilateral shape requires an amplification of non-linear effects around horizon exit and hence modifications to single-field inflation \citep{Chen2007}. Particular examples are non-canonical kinetic terms as in the Dirac-Born-Infeld model \citep{Alishahiha2004} or higher-derivative terms such as in K-inflation \citep{kinflation1999}, ghost inflation \citep{Arkani-Hamed2004}, effective field theories of inflation \citep{Cheung2008} or Galileon inflation \citep{Burrage2011}. The orthogonal shape \citep{Senatore2010} is able to distinguish between variants of non-canonical  kinetic terms and higher-derivative interactions.

The late-time matter density PDF at a given smoothing scale is mostly sensitive to the skewness of the primordial density field at that scale and to the running of that skewness around the smoothing scale. As such - unless the PDF is measured on a wide range of smoothing scales - it can only poorly distinguish between different primordial bispectrum shapes. Any model that produces mainly one of the possible bispectrum template can however be successfully tested with PDF measurements. In this paper we consider an analysis of the PDF at redshift $z=1$ in a survey volume of $V=100$ (Gpc$/h$)$^3$, which is smaller than the effective volume of upcomming surveys such as Spherex with $V_{\mathrm{eff}}\approx 150$ (Gpc$/h$)$^3$ and somewhat larger than existing surveys such as BOSS with $V_{\mathrm{eff}}\approx 55$ (Gpc$/h$)$^3$ \citep{Dore2014, Alam2017}. At a smoothing scale of $30$Mpc/$h$ we find our PDF model to agree with the high-resolution run of the Quijote N-body simulations \citep{Navarro2019} to $\lesssim 0.2\%$ accuracy over a range of $\delta[30\mathrm{Mpc}/h] \in [-0.3, 0.4]$. This is within cosmic variance of the considered volume of $100$ (Gpc$/h$)$^3$ (which is also the combined volume of the Quijote high-resolution boxes). Restricting to this smoothing scale and to these mildly non-linear densities we find that a PDF based analysis can measure the amplitude of different primordial bispectrum shapes to an accuracy of \smash{$\Delta f_{\mathrm{NL}}^{\mathrm{loc}} = \pm 7.4\ ,\ \Delta f_{\mathrm{NL}}^{\mathrm{equi}} = \pm 22.0\ ,\ \Delta f_{\mathrm{NL}}^{\mathrm{ortho}} = \pm 46.0\ $} - even when marginalising over the non-linear variance of the density field as a free parameter. When pushing to smaller scales and assuming a joint analysis of the PDF with smoothing radii of $30$Mpc/$h$ and $15$Mpc/$h$ ($\delta[15\mathrm{Mpc}/h] \in [-0.4, 0.5]$) this improves to \smash{$\Delta f_{\mathrm{NL}}^{\mathrm{loc}} = \pm 3.3\ ,\ \Delta f_{\mathrm{NL}}^{\mathrm{equi}} = \pm 11.0\ ,\ \Delta f_{\mathrm{NL}}^{\mathrm{ortho}} = \pm 17.0\ $} - even when marginalizing over the non-linear variances at both scales as two free parameters. Especially, such an analysis can simultaneously measure $f_{\mathrm{NL}}$ and the amplitude and slope of the non-linear power spectrum. Note that any dependence of these forecasts on $\sigma_8$ is completely mitigated by this marginalisation. We do not consider the impact of $\Omega_m$ on our signals \citep[see][for an investigation of the general cosmology dependence of the PDF]{Uhlemann2020} though \citet{Friedrich2018} and \citet{Gruen2018} have demonstrated that parameters of the $\Lambda$CDM model and higher order moments of the density field can be measured simultaneously from what they call lensing-around-cells. Ultimately, we are working towards a combination of a late-time PDF analysis with the early-universe results of \citetalias{Planck2018PNG}. These two analyses have the potential to complement each other: the CMB providing information about the background $\Lambda$CDM spacetime, the late-time density PDF providing information about non-linear structure growth and both of them containing independent information about the imprint of primordial non-Gaussianities on the large scale structure.

Our paper is outlined as follows: \Secref{summary} summarizes our procedure of modelling the matter density PDF and its moments and compares their statistical power for measuring primordial non-Gaussianity.  \Secref{heuristic} provides intuitive explanations for the impact of primordial non-Gaussianity on the density PDF and its moments, while \Secref{model} presents a detailed derivation of our actual modelling approach. \Secref{covariance} describes the different simulations we used and how we estimate the covariance matrix of the density PDF and moments at the scales under consideration. We conclude and discuss our results in \secref{discussion}. \verb|Python| and \verb|C++| tools to carry out the calculations presented in this paper are publicly available\footnote{\url{https://github.com/OliverFHD/CosMomentum}}.

\begin{figure*}
  \includegraphics[width=0.7\textwidth]{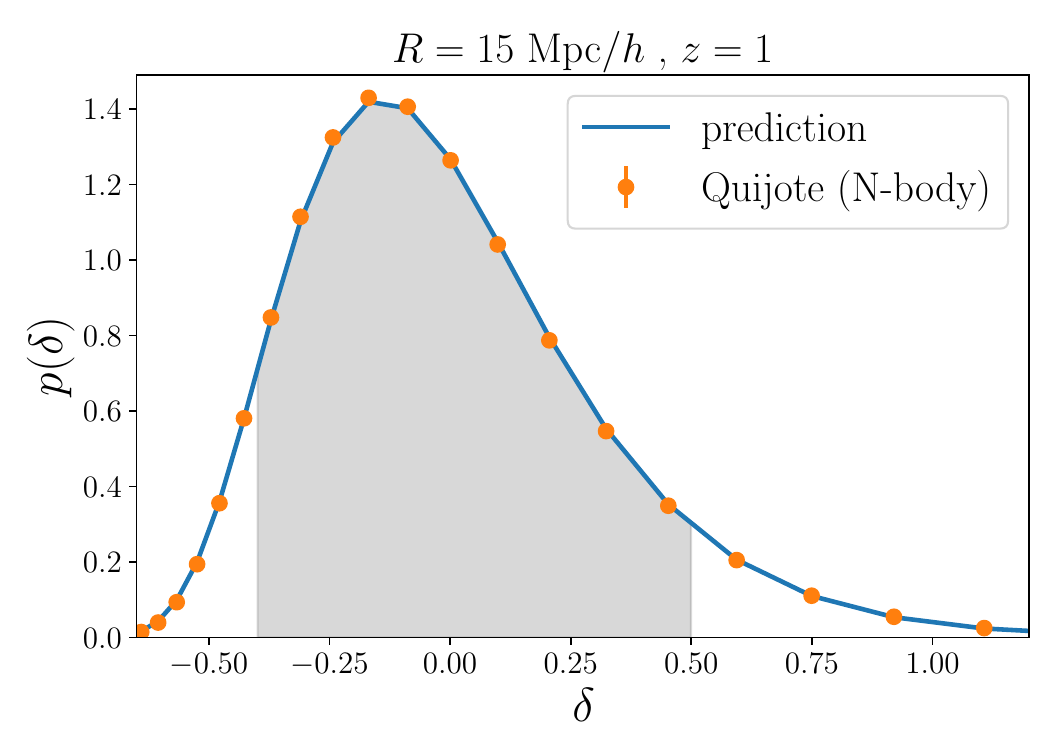}
   \caption{Comparing the matter density PDF measured in the Quijote N-body simulations \citep{Navarro2019} to our analytic model for Gaussian initial conditions. In this paper we compare cosmological information obtained from the bulk of the PDF (grey area, $\approx 87\%$ of probability) to that obtained from moments of the density field. The latter can strongly depend on the tails of the PDF which are impacted more severely by the non-linear evolution of the density field or baryonic physics. Furthermore, methods to recover the matter density PDF from the galaxy density field \citep{Friedrich2018, Gruen2018} require modelling of non-linear tracer bias which is also more difficult in the tails of the PDF.}
  \label{fi:PDF}
\end{figure*}

\begin{figure*}
  \includegraphics[width=0.47\textwidth]{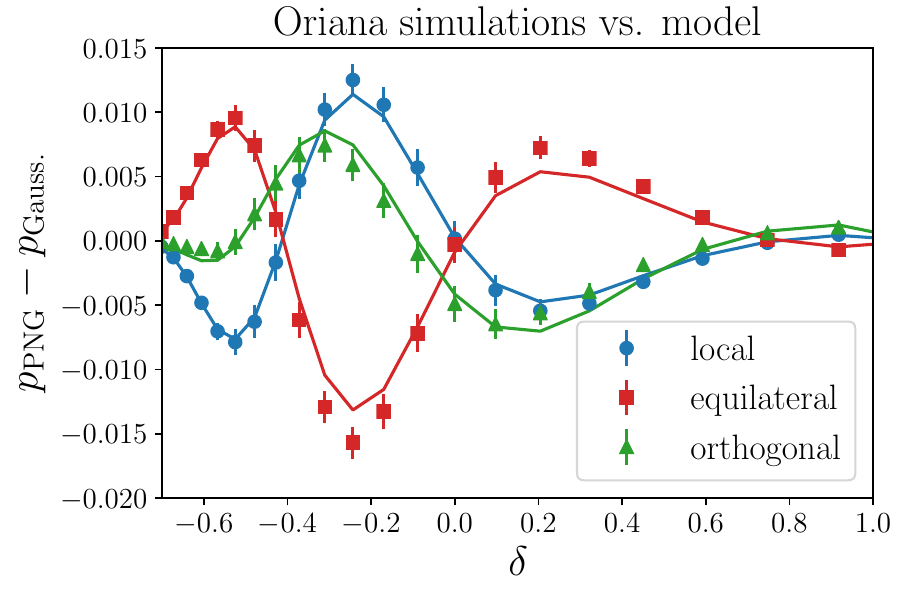}\includegraphics[width=0.47\textwidth]{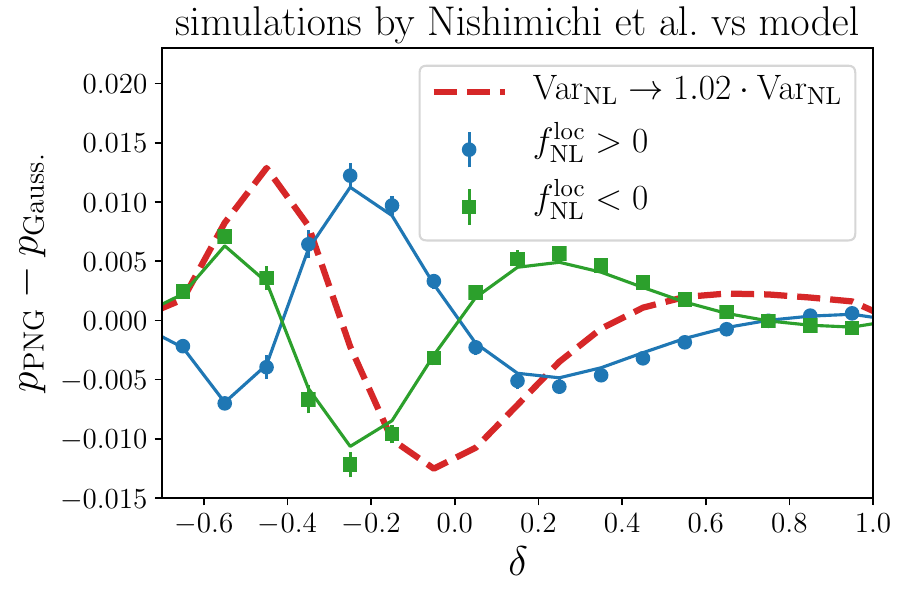}
   \caption{We test the accuracy of our PDF model for exaggerated amplitudes of primordial bispectrum templates at a smoothing scale of $R=15$Mpc$/h$ and redshift $z=1$. \emph{Left:} difference between PDFs obtained from non-Gaussian and Gaussian initial conditions in the Oriana simulations (points with errorbars, see \citealt{Scoccimarro2012}; using $f_{\mathrm{NL}}^{\mathrm{loc}}=100$ $f_{\mathrm{NL}}^{\mathrm{equi}}=-400$ $f_{\mathrm{NL}}^{\mathrm{ortho}}=-400$). Our model predictions for these differences are displayed by the solid lines. \emph{Right:} simulations run by \citet{Nishimichi2012, Valageas2011} for local primordial non-Gaussianity with $f_{\mathrm{NL}}^{\mathrm{loc}}=\pm100$. Note that in all simulations the primordial non-Gaussianity also changes the late-time non-linear variance. We absorb this by fitting different values for this variance to each simulation. The errorbars in each panel are for individual simulations, corresponding to a volume of $~14$ (Gpc$/h$)$^3$ for Oriana and $~70$ (Gpc$/h$)$^3$ for Nishimichi. This actually overestimates the uncertainty since Gaussian and pNG versions of each simulation have strongly correlated initial conditions. The remaining mismatch between model and simulations may seem small, but it is not negligible \wrt the precision of future surveys. In \secref{discussion} we discuss possible causes of these discrepancies and how to address them in the future.}
  \label{fi:PDF_residuals_f_NL_100}
\end{figure*}

\section{Summary of procedure and forecast of statistical power}
\label{sec:summary}

\subsection{Modelling the matter PDF}

\begin{figure*}
  \includegraphics[width=0.49\textwidth]{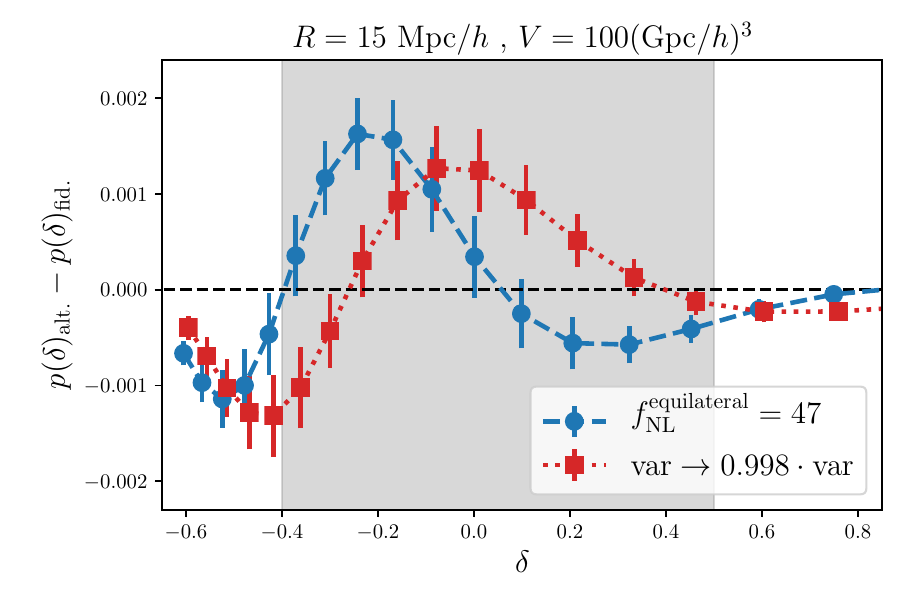}\includegraphics[width=0.49\textwidth]{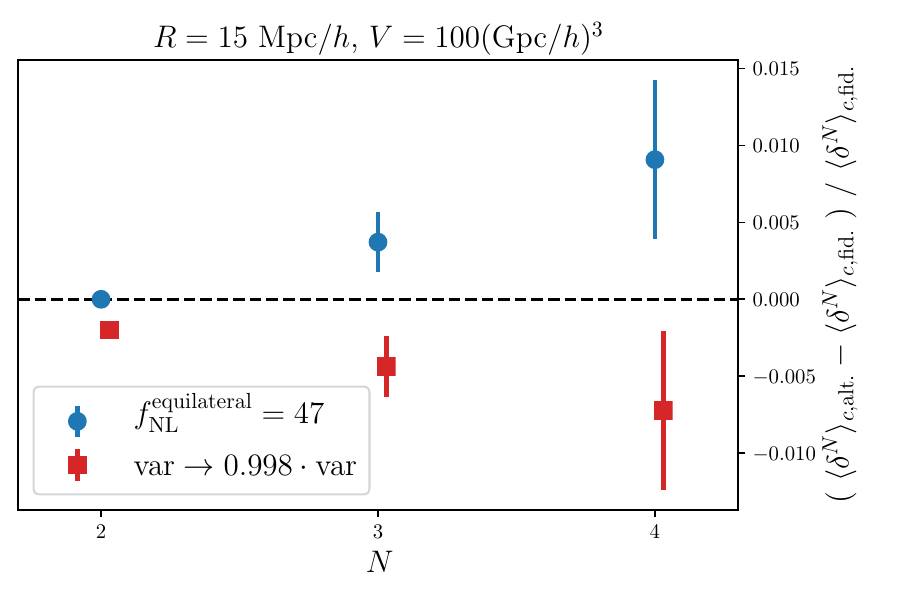}
  
  \includegraphics[width=0.49\textwidth]{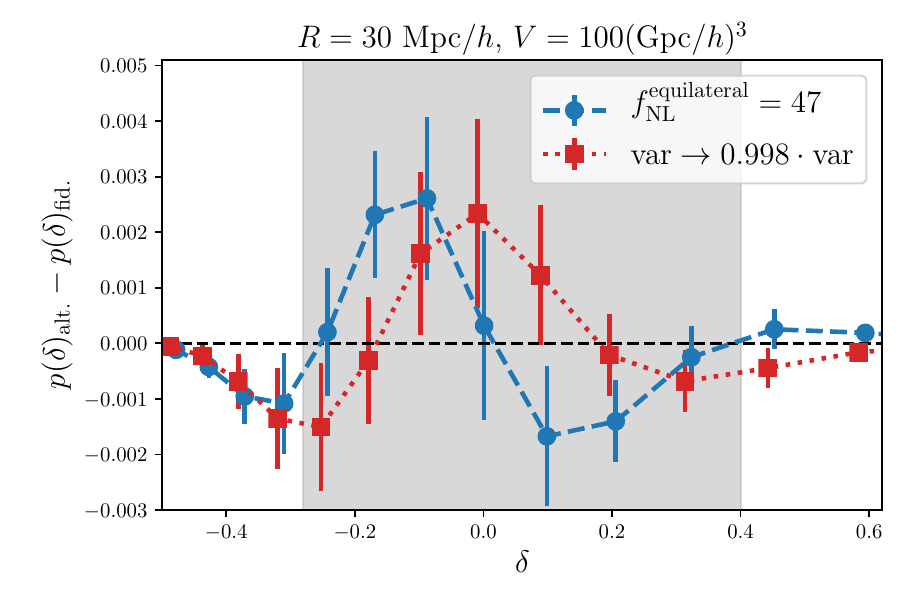}\includegraphics[width=0.49\textwidth]{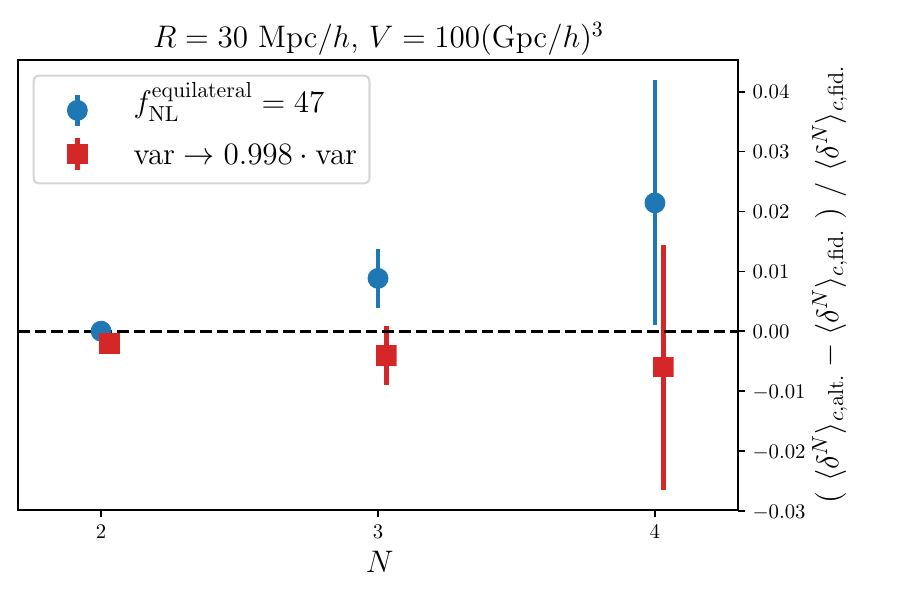}
   \caption{Comparing the predicted response of the PDF (left) and the cumulants (right) of the density field to changes in $f_{\mathrm{NL}}^{\mathrm{equi}}$ (the amplitude of an equilateral shape of the primordial Bispectrum) and in the late-time non-linear variance. In the left panels we show the absolute difference between modified and fiducial PDFs, while on the right panels we show relative differences between modified and fiducial cumulants. The value of \smash{$f_{\mathrm{NL}}^{\mathrm{equi}}=47$} corresponds to the $1\sigma$ uncertainty of \citetalias{Planck2018PNG} (though the latter simultaneously vary both $f_{\mathrm{NL}}^{\mathrm{ortho}}$ and $f_{\mathrm{NL}}^{\mathrm{equi}}$, see our discussion in \apprefnospace{vary_different_shapes}). Error bars assume a survey volume of $V=100(\mathrm{Gpc}/h)^3$. The grey regions indicate the bulk of the PDFs that is used for the forecasts in this work. In total it excludes about $13\%$ of the probability in the tails for $R=15\mathrm{Mpc}/h$ and about $5\%$ of the probability for $R=30\mathrm{Mpc}/h$.} 
  \label{fi:moment_residuals_summary}
\end{figure*}

We start by summarizing the main technical result of this paper. Assume that we know the cumulant generating function (CGF) of the linear density contrast
\begin{equation}
    \varphi_{L, R}(j) \equiv \sum_{n=2}^\infty\ \langle \delta_{L,R}^n \rangle_c\ \frac{j^n}{n!}\ ,
\end{equation}
where $\langle \delta_{L,R}^n \rangle_c$ are the local connected moments (or cumulants) of the linear density contrast field $\delta_L(\mathbf{x})$ today, averaged over spheres of radius $R$
\begin{equation}
\label{eq:defsmooth}
    \delta_{L,R} = \int \text{d}^3x\, W_R(x)\delta_L(\mathbf{x})\ ,\ W_R(x) =\frac{3\Theta(R-|x|)}{4\pi (R)^3} \,.
\end{equation}
For Gaussian initial conditions the linear CGF is simply
\begin{equation}
    \varphi_{L, R}^{\mathrm{Gauss}}(j) =  \frac{\langle \delta_{L,R}^2 \rangle_c}{2}\ j^2\ ,
\end{equation}
while for small primordial non-Gaussianity it can be approximated as
\begin{equation}
    \varphi_{L, R}^{\mathrm{PNG}}(j) \approx  \frac{\langle \delta_{L,R}^2 \rangle_c}{2}\ j^2 + \frac{\langle \delta_{L,R}^3 \rangle_c}{6}\ j^3\ ,
\end{equation}
where the skewness $\langle \delta_{L,R}^3 \rangle_c$ of the linear density contrast can be calculated from the primordial bispectrum, as described in \secref{bispec} \citep[see also][]{Uhlemann2018b}.

We then derive in this paper that the cumulant generating function $\varphi_R(\lambda, z)$ of the non-linear density contrast at red\-shift $z$ and smoothing scale $R$ can be approximated as
\begin{equation}
\label{eq:final_CGF_approximation}
    \varphi_R(\lambda, z) \approx  - s_{\lambda}(\delta^*, j^*)\ ,
\end{equation}
where $\delta^*$ and $j^*$ minimize the function
\begin{equation}
    s_\lambda(\delta, j)\ =\ -\lambda \mathcal{F}(\delta) + j \delta - \varphi_{L, R(1+\mathcal{F}(\delta))^{1/3}}(j)\ ,
\end{equation}
$\mathcal{F}(\delta) = \mathcal{F}(\delta, z)$ being the function that describes the spherical collapse of a density fluctuation that has linear density contrast $\delta_L = \delta$ today (see \apprefnospace{spherical_collapse}). Hence, we derive an approximation for computing the cumulant generating function of the evolved density field directly from the cumulant generating function of the linear density field.

This result extends the path integral approach of \citet{Valageas2002} and \citet{ Valageas2002III} for Gaussian initial conditions and limited types of primordial non-Gaussianity to general non-Gaussian initial conditions. As for Gaussian initial conditions, the above procedure yields the cumulant generating function at leading order in standard perturbation theory. The accuracy of \eqnref{final_CGF_approximation} can be significantly improved with the re-scaling \citep{Bernardeau2015, Uhlemann2018a, Uhlemann2018b, Uhlemann2018c, Friedrich2018}
\begin{align}
    \varphi_R(\lambda, z) =\ & \frac{\sigma_{L,R}^2(z)}{\sigma_{NL,R}^2(z)}\  \varphi_R^{\mathrm{l.o.}}\left(\lambda\ \frac{\sigma_{NL,R}^2(z)}{\sigma_{L,R}^2(z)}, z\right)\nonumber \\
    =\ & \frac{\sigma_{L,R}^2(z)}{\sigma_{NL,R}^2(z)}\ \sum_{n=2}^\infty\ S_n^{\mathrm{l.o.}}(z)\ \sigma_{NL,R}^{2(n-1)}(z)\ \frac{\lambda^n}{n!}\ .
    \label{eq:CGFrescaling}
\end{align}
Here, $\varphi_R^{\mathrm{l.o.}}$ is the leading order CGF from \eqnrefnospace{final_CGF_approximation}, $S_n$ are the reduced cumulants defined as
\begin{equation}
\label{eq:reduced_cumulants_definition}
    S_n\ \equiv\ \frac{\langle \delta_R^n \rangle_c}{\sigma_{R}^{2(n-1)}}\,,
\end{equation}
and $\sigma_{L,R}^2(z)$ and $\sigma_{NL,R}^2(z)$ are the variances of the linear and non-linear density field at smoothing radius $R$. These variances can be calculated from the linear power spectrum $P_L$ and nonlinear power spectrum $P_{NL}$ as
\begin{equation}
\sigma^2_{L/NL,R}(z) 
= \int \, \frac{\dd k}{2\pi^2}\ P_{L/NL}(k,z)\ k^2\ \tilde W^2_{R}(k)\,,
\label{eq:variance}
\end{equation}
where $\tilde W_{R}(k)$ is the Fourier transform of the spherical top-hat kernel from \eqnref{defsmooth}, given by (\cf \apprefnospace{Fourier_conventions})
\begin{equation}
\label{eq:filter}
W_{R}(k)=3\left(\frac{\sin(Rk)}{(kR)^3} - \frac{\cos(Rk)}{(kR)^2} \right)\ .
\end{equation}
The non-linear power spectrum required for the re-scaling in \eqnref{CGFrescaling} can be obtained from N-body simulations, from fitting formulae such as halofit \citep{Smith2003, Takahashi2012} or from response-function based approaches \citep[\eg respresso, ][]{Nishimichi17}, see \citet{Uhlemann2020} for a comparison. In this work, we treat the nonlinear variance as a free parameter in order to mitigate potential theoretical uncertainty in the modelling of late-time structure growth. Note that marginalising over the amplitude of non-linear density fluctuations makes our $f_{\rm NL}$ constraints also independent of $\sigma_8$ (the amplitude of the linear density contrast field on an 8Mpc$/h$ smoothing scale; see also \citealt{Uhlemann2020} for a discussion of the dependence of the density PDF on $\Lambda$CDM parameters).

Once the cumulant generating function $\varphi_R(\lambda, z)$ has been calculated, the PDF of $\delta_R$ can be obtained from an inverse Laplace transform, \ie
\begin{equation}
\label{eq:PDF_as_Laplace_transform}
p(\delta_R, z) = \int_{-\infty}^{\infty} \frac{\dd \lambda}{2\pi}\ e^{-i\lambda\delta_R + \varphi_R(i\lambda, z)}\ .
\end{equation}
A description of how to efficiently solve this integral is provided in \citet{Valageas2002,Bernardeau2015,Friedrich2018}.

In \figref{PDF} we show a comparison of this model to the density PDF measured in the Quijote N-body simulations \citep[see][details are also given in our \secrefnospace{sims}]{Navarro2019}. The figure shows the PDFs at redshift $z=1$ and for a smoothing radius of $R = 15$Mpc$/h$. \Figref{PDF_residuals_f_NL_100} displays the difference between PDFs evolved from Gaussian initial conditions and PDFs evolved from primordial non-Gaussianity with different primordial Bispectrum shapes. The left panel compares our model to simulations run by \citet{Scoccimarro2012}, while the right panel uses simulated data based on methods developed in \citet{Nishimichi2012, Valageas2011} (see also \citealt{Uhlemann2018b} or our \secrefnospace{sims} for details). It can be seen there that our analytical model realistically captures the shape of the PDF as found in N-body simulations. Note that the approximations made in \secrefnospace{model} become more precise in the limit of $f_{\mathrm{NL}}\rightarrow$ 0 and that also the finite resolution of the simulations influences the comparison of \figrefnospace{PDF_residuals_f_NL_100}, as we discuss in \secrefnospace{discussion}.

\begin{centering}
\begin{figure}
  \includegraphics[width=0.47\textwidth]{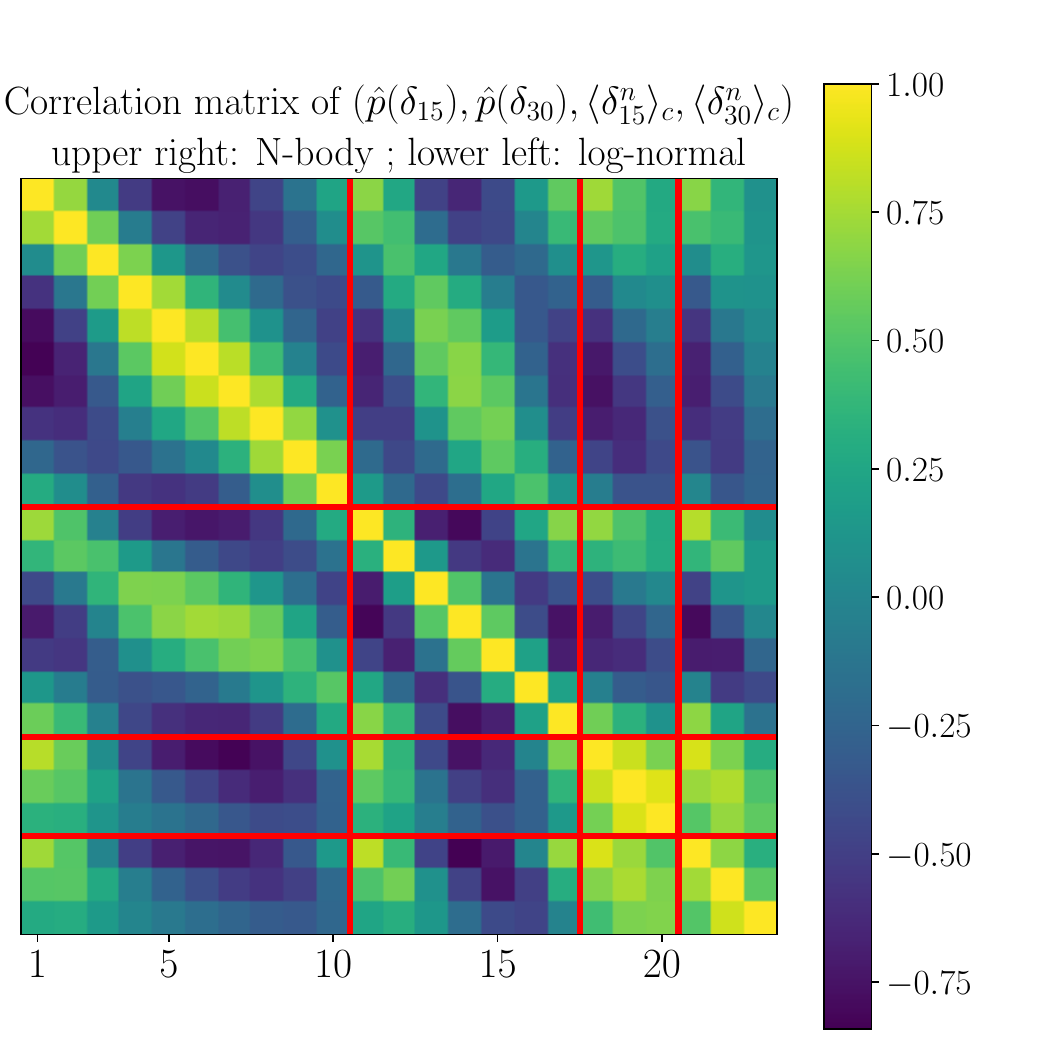}
   \caption{Correlation matrix of the combined data vector of the PDF and cumulants measured at two smoothing scales. The first block (1-10) is the PDF measured on a $15\mathrm{Mpc}/h$ smoothing scale and for density contrasts $\delta \in [-0.4, 0.5]$. The second block (11-17) is the PDF measured on a $30\mathrm{Mpc}/h$ smoothing scale and for density contrasts $\delta \in [-0.3, 0.4]$. The binning of the PDF within these ranges was chosen to match the binning of the fiducial Quijote data products \citep{Navarro2019}. The last two blocks (18-20 and 21-23) are the variance, skewness and kurtosis measured on smoothing scales of $15\mathrm{Mpc}/h$ and $30\mathrm{Mpc}/h$ respectively. The upper right triangle uses the Quijote N-body simulations. This is the covariance that we are using in our forecasts. To investigate a cheap way of producing covariances for future analyses we also investigate log-normal simulations that are tuned to produce the correct variance and skewness on a $15\mathrm{Mpc}/h$ scale. The correlation matrix obtained from these is shown in the lower left triangle (\cf \secrefnospace{lognormal_sims}).}
  \label{fi:correlation_matrix}
\end{figure}
\end{centering}

Since our model captures the impact of primordial non-Gaussianities on the late-time density PDF realistically, we now discuss the impact of values of $f_{\rm NL}$ that are compatible with current experimental bounds. In \figref{moment_residuals_summary} we show the theoretically predicted response of the PDF and its $2$nd, $3$rd and $4$th cumulants (the variance, skewness and kurtosis) to a primordial bispectrum of equilateral shape and with amplitude $f_{\mathrm{NL}}^{\mathrm{equi}}=47$, corresponding to the $1\sigma$ uncertainty of \citetalias{Planck2018PNG}\footnote{Note that while Planck can simultaneously constrain equilateral and orthogonal type non-Gaussianity, the PDF is only sensitive to their combination, as we discuss in \apprefnospace{vary_different_shapes}.}. For these figures, the non-linear variance of the late-time density field was calculated with the halofit power spectrum \citep{Smith2003, Takahashi2012} and the higher order cumulants have been obtained by approximating the cumulant generating function of \eqnref{final_CGF_approximation} with a polynomial (which is nummerically highly non-trivial, please see our discussion in \secrefnospace{numerical_details_moment_extraction}). We also compare this to the response of the PDF when decreasing the non-linear variance by 2 per-mille (which leads to signatures of a similar amplitude). As you can see in the figure, changes in the amplitude of the primordial Bispectrum and changes in the late-time variance have non-degenerate signatures on the shape of the PDF. The errorbars shown in the figure represent cosmic variance for a survey volume of $V=100$(Gpc$/h$)$^3$ at at redshift $z=1$. This corresponds to the combined volume of the high-resolution runs of Quijote which is smaller than the effective volume of upcomming surveys such as Spherex with $V_{\mathrm{eff}}\approx 150$ (Gpc$/h$)$^3$ but somewhat larger than existing surveys such as BOSS with $V_{\mathrm{eff}}\approx 55$ (Gpc$/h$)$^3$ \citep{Dore2014, Alam2017}. Our error bars are obtained - as part of the full covariance matrix of PDFs and moments at the two radii $R=15$Mpc$/h$ and $R=30$Mpc$/h$ - from the fiducial Quijote runs (see \secref{covariance} for details and \figref{correlation_matrix} for a display of the full correlation matrix). An important point to note here is: While the agreement between our model and N-body simulations is at the sub-percent level for the total PDF (\cf \figrefnospace{Quijote_residuals}), the enormous statistical power of future surveys such as Spherex will require per-mille level accuracy. This will require careful control of both theoretical errors for the predictions and finite resolution effects in the simulations, as we discuss in \secref{discussion}.

\begin{figure}
  \includegraphics[width=0.47\textwidth]{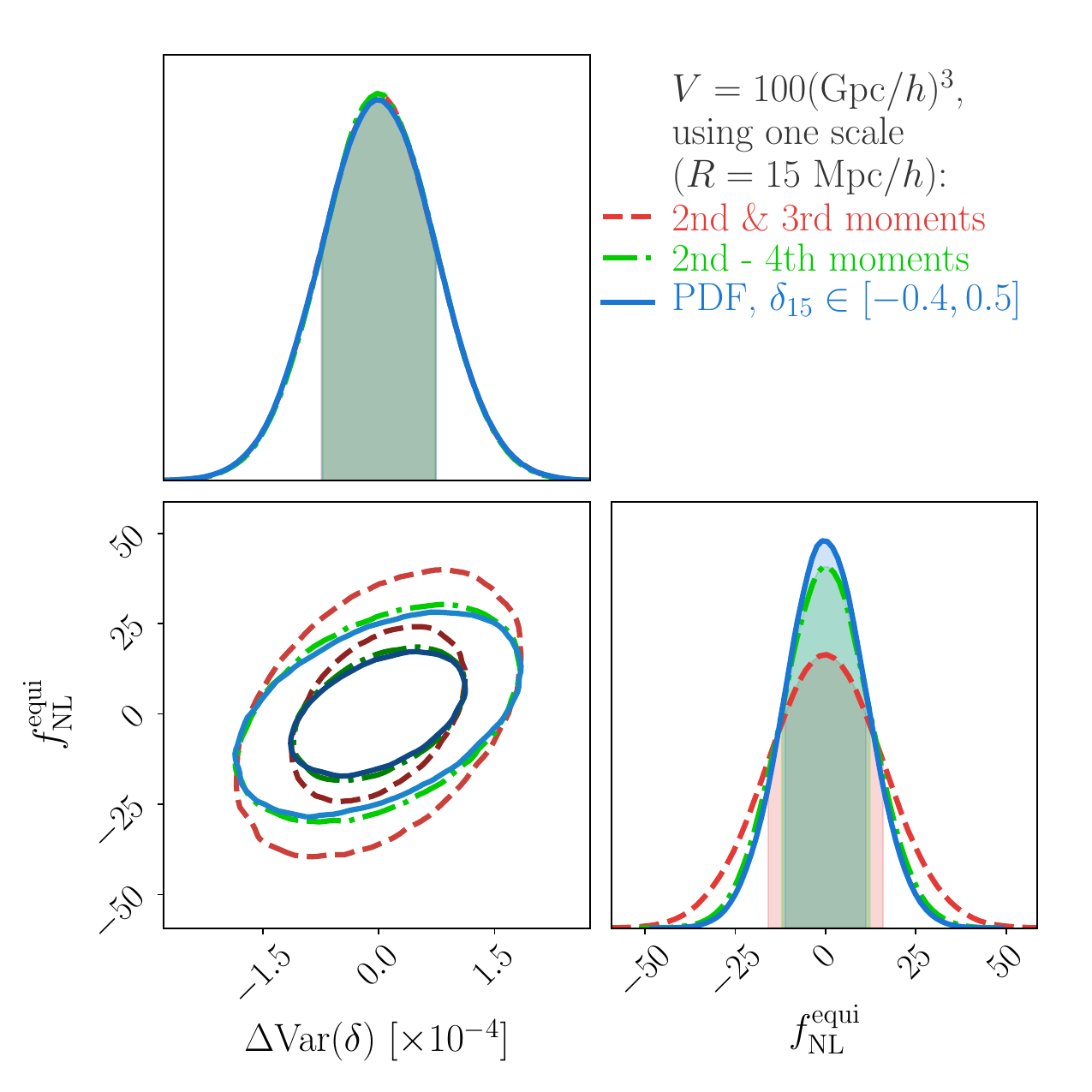}
  
  \includegraphics[width=0.47\textwidth]{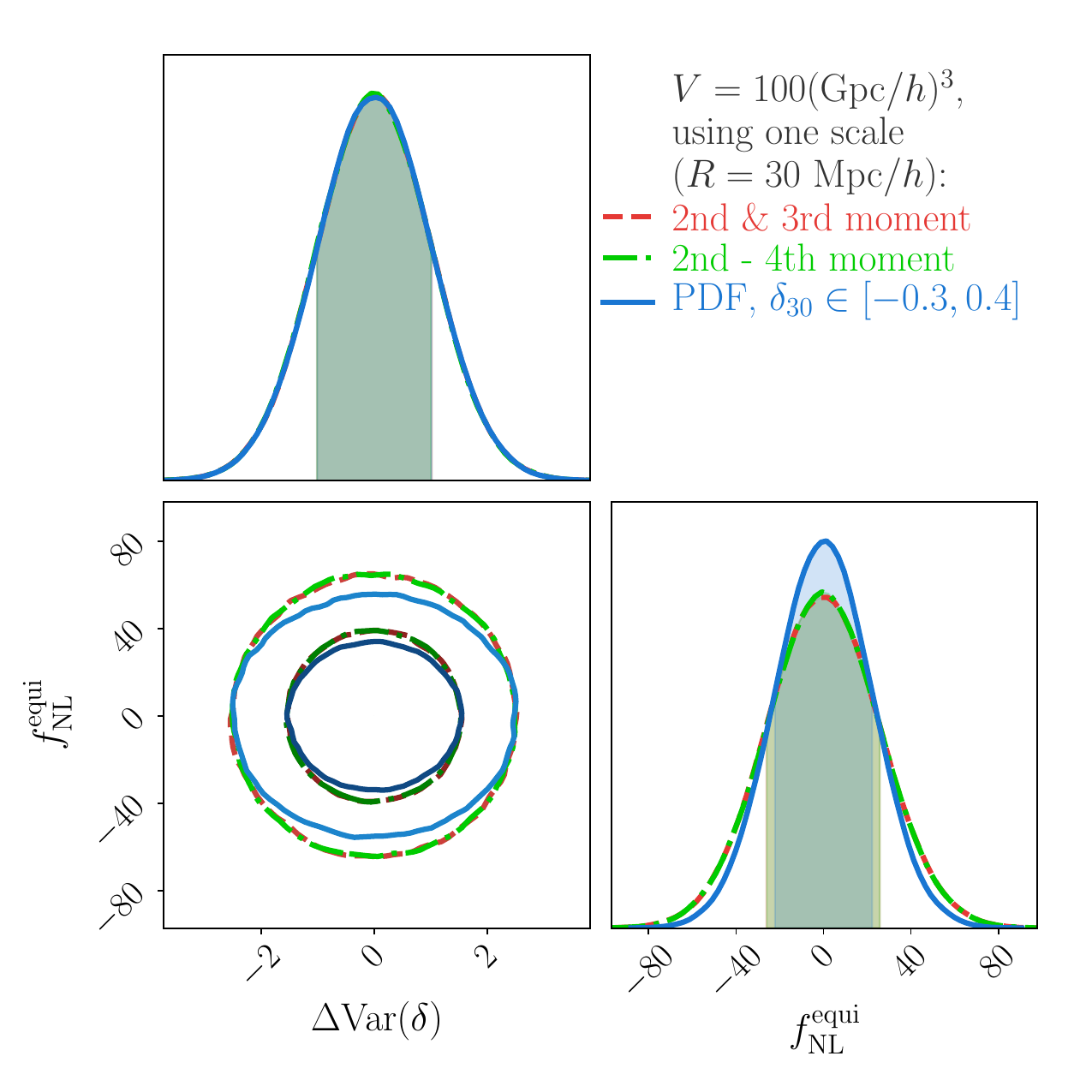}
   \caption{Same as the upper panel of \figref{constraints_joint} but when analysing the PDF or cumulants individually at the smoothing scales $R=15$Mpc$/h$ (top) and $R=30$Mpc$/h$ (bottom). The smaller scale of $R=15$Mpc$/h$ seems to provide more power to constrain primordial non-Gaussianity. For the larger smoothing scale of $R=30$Mpc$/h$ the information obtained from a moment-based analysis seems to saturate already at the 3rd order cumulant (the skewness) while for $R=15$Mpc$/h$ the kurtosis still significantly increases the information content. The may be expected since the density field becomes increasingly Gaussian as one moves to larger smoothing scales.}
  \label{fi:constraints_R15}
\end{figure}

\subsection{Constraining $f_{\rm NL}$ with the PDF and cumulants}

\begin{figure*}
\begin{center}
  \includegraphics[width=0.75\textwidth]{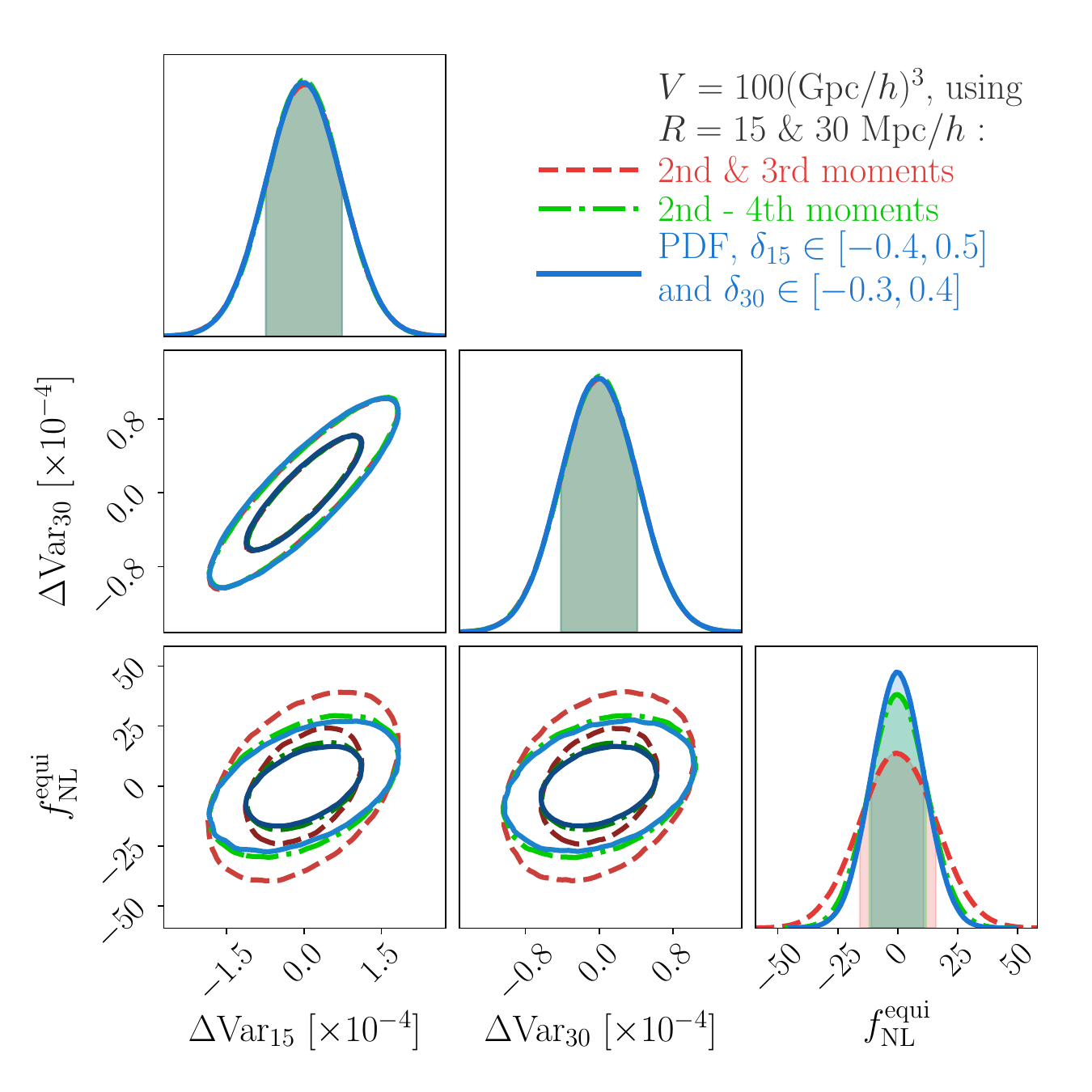}
  \end{center}
   \caption{Forecast for measurements of $f_{\mathrm{NL}}^{\mathrm{equi}}$ (the amplitude of a primordial Bispectrum with equilateral shape, \citealt{Fergusson2009}) and the late-time variance of the density field when analysing PDF or cumulants at both $R=15$Mpc$/h$ and $R=30$Mpc$/h$. Note that we are able to simultaneously measure the amplitude of the primordial Bispectrum and both the amplitude and slope of the late-time power spectrum. Marginalisation over the late-time variances also makes our constraints on primordial non-Gaussianity insensitive to $\sigma_8$. Ultimately we are preparing a joint analysis of the late-time density PDF and early-universe constraints from \citetalias{Planck2018PNG}, which is why we keep all other cosmological parameters fixed in this forecast \citep[see also][for an investigation of the general cosmology dependence of the PDF]{Uhlemann2020}.}
  \label{fi:constraints_joint}
\end{figure*}

Based on our theoretical model and the covariance matrix estimated from the Quijote simulations we investigate how well analyses of the matter density PDF can measure the amplitude of the different primordial bispectrum templates as well as the late-time amplitude of density fluctuations. To do so, we consider the 3 parameters $\boldsymbol{\theta} = [\mathrm{Var}_{15}, \mathrm{Var}_{30}, f_{\mathrm{NL}}]$, where $\mathrm{Var}_{R}$ denotes the non-linear variance of the density contrast at smoothing scale $R$ and $z=1$ and $f_{\mathrm{NL}}$ denotes the amplitude of different primordial bispectrum templates. Given a model for a data vector, $\mathbf{x} = \mathbf{x}[\boldsymbol{\theta}]$, and an expected covariance matrix $\mathbf{C}$ for that data vector, one can estimate the covariance matrix $\mathbf{C}_{\mathrm{param}}$ of the statistical uncertainties in the parameters as \citep{Krause2017}
\begin{equation}
\label{eq:Fisher}
    \left(\mathbf{C}_{\mathrm{param}}^{-1}\right)_{ij} = \frac{\dd \mathbf{x}}{\dd \theta_i}\cdot \mathbf{C}^{-1}\cdot\frac{\dd \mathbf{x}}{\dd \theta_j}\ .
\end{equation}
This assumes that the noise in measurements $\mathbf{\hat x}$ of $\mathbf{x}$ has a multivariate Gaussian distribution and that the dependence of $\mathbf{x}$ on the parameters $\boldsymbol{\theta}$ is close to linear. In our case, $\mathbf{x}$ is either of the following data vectors:
\begin{itemize}
    \item the PDF measured for $\delta_{15\mathrm{Mpc}/h} \in [-0.4, 0.5]$ ($\approx 87\%$ of probability) or $\delta_{30\mathrm{Mpc}/h} \in [-0.3, 0.4]$ ($\approx 95\%$ of probability), see the blue contours in \figref{constraints_R15}. 
    \item measurements of the first two non-vanishing cumulants (variance and skewness) or the first three non-zero cumulants (variance, skewness and kurtosis) of the density field at these two smoothing scales, see the red and green contours in \figref{constraints_R15}. 
    \item the combined data vector of either the PDF or the cumulants measured at both smoothing scales, see \figref{constraints_joint}. 
\end{itemize}
The above cuts in the PDFs where chosen such that they remove approximately the same amount of probability in both the underdense and overdense tails. The range $\delta_{15\mathrm{Mpc}/h} \in [-0.4, 0.5]$ is motivated by demanding that our PDF model be in $\sim 1\%$ agreement with the high-resolution runs of the Quijote simulations. The motivation for choosing the range $\delta_{30\mathrm{Mpc}/h} \in [-0.3, 0.4]$ was to cut enough of the tail probabilities in order to assume multivariate Gaussian noise on the PDF measurements (see explanations below).

We estimate the covariance of each of these data vectors from the Quijote simulations and choose our binning of the PDFs to match that of \citet{Navarro2019}. Using the ensemble of PDF and cumulant measurements from Quijote we also test that individual data points have a close to Gaussian distribution. If we were to analyse the PDFs over the entire range $\delta_R \in [-1, \infty]$, then the noise of PDF measurements could not have a multivariate Gaussian distribution, because of the normalization condition$\int p(\delta_R) \dd \delta_R=1$. Also, PDF measurements will always be positive which necessarily skews their distribution. This is especially noticeable in the tails of the PDF, where sampling noise is expected to lead to a Poisson-like rather than a Gaussian noise. Both of these problems are alleviated in our analysis because we only consider the bulk of the PDF. We investigate multivariate statistical behavior of the PDF measurements in the Quijote simulations in \appref{Gaussian_likelihood} and find that it is indeed well described by a multivariate Gaussian distribution.

\Figref{constraints_R15} shows the constraints on the amplitude of an equilateral primordial bispectrum, $f_{\mathrm{NL}}^{\mathrm{equi}}$ that can be obtained when analysing the PDF at either of the smoothing radii $R=15$Mpc$/h$ and $R=30$Mpc$/h$ - again assuming a survey volume of $V=100(\mathrm{Gpc}/h)^3$ at $z=1$ (blue solid contours). A comparison of the upper and lower panel of the figure indicates that the smaller scale is more powerful in constraining $f_{\mathrm{NL}}$. We want to stress again that our model is only accurate enough to analyse the PDF at the $30$Mpc$/h$ smoothing scale. At $R=15$Mpc$/h$ the residual $\lesssim 0.8\%$ modelling error is significantly larger than the cosmic variance of the considered survey volume, \cf\ the discussion in \secrefnospace{model_precision}. As explained in that section, our remaining inaccuracy likely result from next-to-leading order corrections derived in \citet{Ivanov2019}. These corrections should be included in any real data analysis but we do not expect them to significantly impact the Fisher analysis presented here.

In both panels of \figref{constraints_R15} we also show the constraints on $f_{\mathrm{NL}}^{\mathrm{equi}}$ that can be obtained from direct measurements of the variance and skewness of density fluctuations at these scales (red dash-dotted contours) and from the variance, skewness and kurtosis combined (green dashed contours). In can be seen there, that the PDF indeed contains more information than just the 2nd and 3rd moment of fluctuations combined. The moment-based analysis only catches up with the PDF-based one once the kurtosis is also considered. At the same time, smoothing scale including the kurtosis in the moment based analysis doesn't add as much information for $R=30$Mpc$/h$ as it does for the $15$Mpc$/h$ smoothing scale. This may be expected because the density field becomes increasingly Gaussian at larger smoothing scales.

In \figref{constraints_joint} we show the statistical power achievable with a joint analysis of the PDF at smoothing radii $R=15$Mpc$/h$ and $R=30$Mpc$/h$ - again assuming a survey volume of $V=100(\mathrm{Gpc}/h)^3$ at $z=1$ (blue solid contours). Also, we show the constraints obtainable from analyses of the variance and skewness (red dash-dotted contour) and the variance, skewness and kurtosis (green dashed contour). Note again that in each of these cases we consider the non-linear variance of density fluctuations at both smoothing scales as two free parameters. This means that the analyses tested here can simultaneously measure primordial non-Gaussianity and the amplitude and slope of the non-linear power spectrum.

\begin{table}
\begin{tabular}{c|ccc}
analysis & $\Delta f_{\mathrm{NL}}^{\mathrm{local}}$ & $\Delta f_{\mathrm{NL}}^{\mathrm{equi}}$ & $\Delta f_{\mathrm{NL}}^{\mathrm{ortho}}$\\
\hline
PDF ($R=30$Mpc$/h$) & $\pm 7.3$ & $\pm 22$ & $\pm 46$\\
2 moments ($R=30$Mpc$/h$) & $\pm 8.5$ & $\pm 26$ & $\pm 54$\\
3 moments ($R=30$Mpc$/h$) & $\pm 8.5$ & $\pm 26$ & $\pm 54$\\
\hline
PDF ($R=15$Mpc$/h$) & $\pm 3.4$ & $\pm 11.2$ & $\pm 18$\\
2 moments ($R=15$Mpc$/h$) & $\pm 5.0$ & $\pm 15.8$ & $\pm 33$\\
3 moments ($R=15$Mpc$/h$) & $\pm 3.8$ & $\pm 12.0$ & $\pm 24$\\
\hline
PDF (joint) & $\pm 3.2$ & $\pm 10.8$ & $\pm 17$\\
2 moments (joint) & $\pm 5.0$ & $\pm 15.7$ & $\pm 33$\\
3 moments (joint) & $\pm 3.7$ & $\pm 11.8$ & $\pm 24$\\
\end{tabular}
\caption{Forecast for measurements of the amplitude of three different types of primordial bispectra from a PDF-based and moment-based analyses. The constraints are marginalised over the late-time variances $\mathrm{Var}_{15}$ and $\mathrm{Var}_{30}$ as two independent nuisance parameters. Note however, that we do NOT simultaneously vary the amplitude of different bispectrum shapes. In circular apertures, both moments and PDF can hardly distinguish between them (\cf \apprefnospace{vary_different_shapes}). Note also, that further modelling improvements are needed to actually make use of the $15$Mpc$/h$ scale since on that scale our model only agrees with N-body results to $\sim 0.8\%$ (\cf the discussion in \secref{model_precision}).}
\label{tab:Fisher_summary}
\end{table}

In \tabref{Fisher_summary} we summarize the constraints of the considered analyses for different bispectrum types \citep{Fergusson2009, Scoccimarro2012} - the equilateral template ($f_{\mathrm{NL}}^{\mathrm{equi}}$), the orthogonal template ($f_{\mathrm{NL}}^{\mathrm{ortho}}$) and a bispectrum from local non-Gaussianity ($f_{\mathrm{NL}}^{\mathrm{local}}$). For the smoothing scale of $30$Mpc$/h$ these constraints are compatible with current measurements of $f_{\mathrm{NL}}$ from the CMB \citetalias{Planck2018PNG}. If our model accuracy can be improved along the lines of \citet{Ivanov2019} to also encompass the $15$Mpc$/h$ scale then the PDF-based analysis could even significantly improve over the CMB measurements. (As could the moment based analysis. However, we have not rigorously evaluated the accuracy of our moment predictions. Especially, we expect them to suffer more from modelling uncertainties in the high-density tail of the PDF.) We want to stress that the PDF alone can not distinguish between different bispectrum shapes, as we discuss in \apprefnospace{vary_different_shapes}. Hence, we have to consider one bispectrum template at a time in \figref{constraints_joint}. Eventually, we are working towards combining an analysis of the late-time PDF with the early-universe results of \citetalias{Planck2018PNG}. In such a combined analysis, the CMB would provide information about the background $\Lambda$CDM spacetime (whose parameters are largely fixed here) while the late-time density PDF will constrain non-linear structure growth and the imprint of primordial non-Gaussianity on the late-time large scale structure.

We explain the details of these results in \secref{model} and \secref{covariance}.

\section{Gaining intuition for the impact of primordial non-Gaussianity}
\label{sec:heuristic}

Before presenting the details of our full modelling approach in \secref{model} we want to provide approximations that make the imprints of non-Gaussian initial conditions, encoded in that modelling approach, more transparent.

\subsection{Impact of primordial skewness on cumulants}
\label{sec:moments}

For Gaussian initial conditions, the reduced cumulants $\langle \delta^n \rangle_c/\langle \delta^2 \rangle_c^{n-1}$ of the unsmoothed density contrast are given by simple constants and a smoothing scale $R$ induces only a mild running due to the changing slope in the linear variance  \citep{Peebles1980,Bernardeau1994}. At leading order this gives 
\begin{align}
\label{eq:S34G}
    S_3^{\mathrm{G}} &= \frac{34}{7} + \gamma\ ,\quad \gamma=\frac{d\log \sigma^2_L(R)}{d\log R}\,,\\
    S_4^{\mathrm{G}} &= \frac{60712}{1323} + \frac{62}{3} \gamma + \frac{7}{3}\gamma^2 \, \nonumber
\end{align}
for the 3rd and 4th order reduced cumulants. For primordial non-Gaussianity, analytical predictions for the reduced skewness without smoothing by \citet{Fry1994} demonstrate that the Gaussian result is offset by terms depending on the initial 3-point correlation function (and higher-point correlations that we neglect here). \citet{Chodorowski1996} generalised this result to the unsmoothed kurtosis, which is also coupled to the initial skewness by non-linear evolution. \citet{Gaztanaga1998} used spherical collapse to relate the reduced cumulants $S_N^{\rm NG}$ in the presence of primordial non-Gaussianity to their Gaussian counterparts from equation~\eqref{eq:S34G}. Denoting the 3rd central moment of the linear density field as $\kappa_3^L$, they obtain
\begin{align}
\label{eq:S34NG}
    S_3^{\rm NG} &= S_3^{\rm G} + \frac{1}{\sigma_L}\frac{\kappa_3^L}{\sigma_L^3} - 2 \left(\frac{S_3^{\rm G}}{3}-1\right) \left(\frac{\kappa_3^L}{\sigma_L^3}\right)^2\\
    S_4^{\rm NG} &= S_4^{\rm G} + \frac{4 S_3^{\rm G}}{\sigma_L} \frac{\kappa_3^L}{\sigma_L^3} + \left(3+7 S_3^{\rm G} - \frac{14}{3} (S_3^{\rm G})^2 + \frac{3}{2} S_4^{\rm G}\right) \frac{\kappa_3^L}{\sigma_L^3} \nonumber \,.
\end{align}
The most important contribution to $S_3$ is given in terms of the linearly evolved reduced skewness (induced by primordial non-Gaussianity), $\kappa_3^L/\sigma_L^4$, which decays linearly with the growth function, because $\delta_L(z)\propto D(z)$. On the other hand, nonlinear evolution leads to almost constant reduced cumulants $S_N^{\rm G}$, such that the primordial skewness offsets the nonlinear reduced skewness and kurtosis by an amount that is inversely proportional to the growth function. Note that this signature is qualitatively different from the impact of $\Lambda$CDM parameters, which change the linear variance in \eqnref{S34G}, but leave the hierarchical ratios $S_N$ close to constant in time \citep[for a discussion, see Section 2.1 in][]{Uhlemann2020}. In a multi-redshift analysis, we expect that this property would allow to jointly constrain $f_{\rm NL}$ as amplitude of the linear skewness, and $\Lambda$CDM parameters like $\Omega_m$ and $n_s$ which drive the scale-dependence of the linear variance.

Let us now quantitatively estimate the effect of different primordial bispectrum shapes discussed in \secref{bispec}. The amplitudes of the initial rescaled skewness for a radius of $R=15$ Mpc$/h$ for different templates are
\begin{equation}
    \frac{\kappa^L_3}{\sigma_L^3} \left(15 \text{Mpc}/h\right)\simeq \{28 f_{\rm NL}^{\rm loc},8.5 f_{\rm NL}^{\rm equi},-5 f_{\rm NL}^{\rm orth}\}\cdot 10^{-5}\,.
\end{equation}
The predicted ratios of reduced cumulants from Equations~\ref{eq:S34G}-\ref{eq:S34NG} agree well with the measurements in the Oriana simulations for radii $R=20-40$ Mpc$/h$ at redshift $z=0.34$ \citep[see Figure 1 in][]{Mao2014}.
Since the nonlinear variance depends very weakly on the amplitude of primordial non-Gaussianity $f_{\rm NL}$ \citep{Uhlemann2018b}, the ratio of connected moments is close to the ratio of reduced cumulants from \eqnref{defSn}. 
When considering the equilateral model with $f_{\rm NL}^{\rm equi}=47$, we obtain a $0.4 \%$ increase for the skewness and a $1\%$ increase in the kurtosis, in good agreement with the result from the full shape of the matter PDF shown in the upper right panel of \figref{moment_residuals_summary}.

\subsection{Heuristic approximation for the matter PDF}
While we use the recipe described in Section~\ref{sec:recipe} to model the density PDF, one can gain intuition for its sensitivity to primordial non-Gaussianity from a simplistic saddle-point approximation for the PDF from \eqnref{PDF_as_Laplace_transform} \citep[see][]{Uhlemann2018b}. According to the large-deviation principle \citep{BernardeauReimberg2016}, the exponential decay of the late-time density PDF with increasing density contrast, $\psi(\delta_R)$, can be predicted from the cumulant generating function and the reduced cumulants entering in \eqnref{CGFrescaling} are determined by spherical collapse. The matter PDF can then be expressed in terms of this decay-rate function as
\begin{align}
    p(\delta_R,z) & \propto  \sqrt{\psi_{R,z}''(\delta_R)+\frac{\psi_{R,z}'(\delta_R)}{1+\delta_R}}\exp\left[-\psi_{R,z}(\delta_R)\right]\,, \\
    \psi_{R,z}(\delta_R)&\propto \frac{\delta_L^2(\delta_R)}{\sigma_L^2(R_L(R,\delta_R),z)}
    \left[1-\frac{\kappa_3^L(R_L(R,\delta_R),z)\delta_L(\delta_R)}{\sigma_L^4(R_L(R,\delta_R),z)}\right]\ , \nonumber
\end{align}
where $\delta_L(\delta_R)$ is the mapping between linear and nonlinear density contrasts, $R_L(R,\delta_R)=R(1+\delta_R)^{1/3}$ accounts for mass conservation, and $\kappa_3^L/\sigma_L^3\propto f_{\rm NL}$ is the linear rescaled skewness caused by the presence of a primordial bispectrum. This approximation reflects that the leading-order corrections to the reduced cumulants are controlled by the primordial skewness, as we have seen explicitly for the skewness and kurtosis in Equations~\ref{eq:S34NG}. The exponent dominates the behaviour in the tails, where a positive primordial skewness leads to an enhancement of high densities and a suppression of low densities. Around the peak of the PDF, the prefactor becomes relevant and leads to an enhancement in the PDF for moderately underdense spheres and a reduction for moderately overdense spheres, in agreement with the left panels of \figref{moment_residuals_summary}.

\section{Saddle point method for general non-Gaussian initial conditions}
\label{sec:model}

We now present a derivation of our main technical result - the approximation in \eqnref{final_CGF_approximation} for the cumulant generating function (CGF) of the late time density contrast $\delta_R$ for arbitrary non-Gaussian initial conditions and our model for the PDF $p(\delta_R)$ that follows from it.

\subsection{Path integral approach for the cumulant generating function}

The cumulant generating function $\varphi_R(\lambda , z)$ of the non-linear density contrast $\delta_R$ is defined as
\begin{align}
    e^{\varphi_R(\lambda , z)} =\ &  \exp\left(\sum_{n=1}^\infty \frac{\langle \delta_R(\mathbf{x}, z)^n \rangle_c}{n!}\ \lambda^n\right)\nonumber \\
    \equiv\ & \langle e^{\lambda \delta_R} \rangle = \int \dd\delta_R\ p(\delta_R | \mathbf{x}, z)\ e^{\lambda \delta_R}\ .
\end{align}
Here $\langle \delta_R(\mathbf{x}, z)^n \rangle_c $ refers to the $n$th connected moment (or cumulant) of the smoothed nonlinear density contrast $\delta_R(\mathbf{x}, z)$ and the second line is the cumulant expansion theorem \citep{BernardeauReview}. Because of homogeneity and isotropy we will only consider $\delta_R(\mathbf{x}=0, z)$ and hence drop the label $\mathbf{x}$. For  simplicity, we will also suppress the dependence of $\delta_R$ on redshift $z$.

In a procedure similar to that of \citet{Valageas2002}, we can write the above expectation value as a functional integral over all  possible configurations of today's linear density contrast,
\begin{align}
    e^{\varphi_R(\lambda)} =\ & \int \mathcal{D}\delta_L\ \mathcal{P}[\delta_L]\ e^{\lambda \delta_R[\delta_L]}\ ,
\end{align}
where $\mathcal{P}[\delta_L]$ is the probability density functional of the linear density contrast field $\delta_L$ and the non-linear density contrast $\delta_R$ has been expressed as a functional of $\delta_L$. (Note again that we dropped the dependence on $z$ from our notation - this dependence is entirely carried by the functional $\delta_R[\cdot] = \delta_R[\cdot , z]$ since we always consider $\delta_L$ at $z=0$).

For Gaussian initial conditions $\delta_L$ will be a Gaussian random field and the probability density functional $\mathcal{P}[\delta_L]$ can be directly expressed through the linear power spectrum $P_L(k)$. This was done by \citet{Valageas2002} to derive an approximation for the late-time CGF from non-Gaussian initial conditions. \citet{Valageas2002III} also studied a limited set of non-Gaussian initial conditions for which explicit expressions of $\mathcal{P}[\delta_L]$ are available.

Here we extend these studies to general non-Gaussian initial conditions. Defining the linear cumulant generating functional associated with $\mathcal{P}[\delta_L]$ as
\begin{equation}
    \Phi[J_L] = \sum_{n=1}^\infty\ \frac{1}{n!} \int \prod_{i=1}^n\ \dd^3 x_i\ J_L(\mathbf{x}_i)\ \xi_{L,n}(\mathbf{x}_1,\ \dots\ , \mathbf{x}_n)
\end{equation}
we can express $\varphi_R$ as
\begin{align}
    e^{\varphi_R(\lambda)} =\ & \frac{1}{\mathcal{N}} \int \mathcal{D}\delta_L\ \mathcal{D}J_L\ e^{\lambda \delta_R[\delta_L] - iJ_L \cdot \delta_L + \Phi[iJ_L]}\ .
\end{align}
The linear cumulant generating functional $\Phi[J_L]$ encodes the initial conditions. For Gaussian initial fluctuations it can be entirely expressed through the linear two-point correlation function and our results would reduce to those of \citet{Valageas2002}. For the primordial non-Gaussianity models we consider here, it will additionally depend on the primordial three-point function $\xi_{L,3}$ or the corresponding bispectrum, for which we discuss concrete templates in Section~\ref{sec:bispec}.
Comparison with \eqnref{PDF_as_Laplace_transform} shows that the normalisation constant $\mathcal{N}$ is formally given by $|2\pi\mathds{1}|$ , \ie the determinant of $2\pi$ times the unit operator in the space of $J_L$'s and hence infinite. But this normalisation will eventually drop in our calculations. Note also that we introduced the abbreviation
\begin{equation}
    J_L \cdot \delta_L \equiv \int \dd^3 x\ J_L(\mathbf{x})\ \delta_L(\mathbf{x})\ .
\end{equation}
We will calculate the above functional integral with Laplace's method, \ie by approximating the integrand around its maximum with a Gaussian functional, which is also called steepest descent or saddle-point method \citep[\eg][]{Valageas2002, Bernardeau2015, Uhlemann2018a}. This means we define an action $S_\lambda$ as
\begin{equation}
\label{eq:def_of_action}
    S_\lambda[\delta_L, J_L] \equiv - \lambda \delta_R[\delta_L] + iJ_L \cdot \delta_L - \Phi[iJ_L]
\end{equation}
and find the saddle point configurations $\delta_L^*$ and $J_L^*$ that minimize this action in order to approximate
\begin{align}
    e^{\varphi_R(\lambda)} =\ & \frac{1}{\mathcal{N}} \int \mathcal{D}\delta_L\ \mathcal{D}J_L\ e^{-S_\lambda[\delta_L, J_L]}\nonumber \\
    \approx\ & \frac{1}{\mathcal{A}^{1/2}}\  e^{-S_\lambda[\delta_L^*, J_L^*]}\ .
\end{align}
Here $\mathcal{A}$ is the determinant of the Hessian matrix of the functional $S_\lambda$ (now considered as a matrix in the combined space of $\delta_L$'s and $J_L$'s) evaluated at the saddle point configurations $\delta_L^*$ and $J_L^*$. The cumulant generating function of the late-time smoothed density contrast is then given by
\begin{equation}
\label{eq:saddle_point_approximation_for_CGF}
    \varphi_R(\lambda) \approx -S_\lambda[\delta_L^*, J_L^*] - \frac{1}{2} \ln \mathcal{A}\ .
\end{equation}
The second term in this approximation is a next-to-leading order correction to the first term \citep[for Gaussian initial conditions it is equivalent to $1$-loop corrections to the CGF, \cf][]{Valageas2002V, Ivanov2019} and will be neglected in this paper. Instead, we apply a re-scaling of the cumulant generating function that accounts for strong late-time non-linearity by leaving the variance of non-linear density fluctuations as a free parameter. We describe this re-scaling in \secref{non_linear_variance_rescaling} after deriving an explicit expression for the first, leading order term of \eqnref{saddle_point_approximation_for_CGF} in the following section. We would like to stress here: omiting the second term in \eqnref{saddle_point_approximation_for_CGF} and using the re-scaling of \secref{non_linear_variance_rescaling} instead is the biggest conceptual weakness of our analysis. We discuss implications of this for the interpretation of our results and future work in \secrefnospace{model_precision}.

\subsection{Minimizing the action $S_\lambda$}
\label{sec:minimizing_the_action}

In the following, let $\mathpzc{d}F/\mathpzc{d}f(\mathbf{x})$ denote functional derivation of a functional $F$ \wrt the function $f$. To calculate the leading order term in \eqnref{saddle_point_approximation_for_CGF} we have to find configurations $\delta_L^*$ and $J_L^*$ such that
\begin{equation}
    \left.\frac{\mathpzc{d} S_\lambda}{\mathpzc{d} \delta_L(\mathbf{x})}\right|_{\delta_L^*, J_L^*} =\ \ 0\ \ = \left.\frac{\mathpzc{d} S_\lambda}{\mathpzc{d} J_L(\mathbf{x})}\right|_{\delta_L^*, J_L^*}\\
\end{equation}
\begin{align}
\label{eq:minimisation_A}
    \Rightarrow\ iJ_L^*(\mathbf{x}) =\ & \lambda\left.\frac{\mathpzc{d} \delta_R}{\mathpzc{d} \delta_L(\mathbf{x})}\right|_{\delta_L^*}\\
\label{eq:minimisation_B}
    \delta_L^*(\mathbf{x}) =\ & \left.\frac{\mathpzc{d} \Phi}{\mathpzc{d} J_L(\mathbf{x})}\right|_{iJ_L^*}\ .
\end{align}
Because we allow for arbitrary shapes of primordial $N$-point functions in the definition of $\Phi$, it is not obvious that these equations have spherically symmetric solutions. We nevertheless show that a spherically symmetric ansatz $\delta_L^*(\mathbf{x}) = \delta_L^*(x)$, $J_L^*(\mathbf{x}) = J_L^*(x)$ solves Equations \ref{eq:minimisation_A} and \ref{eq:minimisation_B} and argue in the end that the spherically symmetric solution is indeed a global minimum of the action - at least for small deviations from primordial Gaussianity.

With a spherically symmetric ansatz \eqnref{minimisation_A} can be solved \citep[along the lines of][]{Valageas2002} by noting that
\begin{align}
\label{eq:delta_R_equals_F}
    \delta_R[\delta_L^*] =\ & \mathcal{F}(\delta_{L,R_L}^*)\ ,
\end{align}
where the function $\mathcal{F}$ describes spherical collapse of the density fluctuation (\cf \apprefnospace{spherical_collapse}) from the initial, linear radius
\begin{align}
     R_L =\ & R\ (1+\delta_R[\delta_L^*])^{1/3}
\end{align}
to the final radius R. Hence one can see that
\begin{align}
 & \left.\frac{\mathpzc{d} \delta_R}{\mathpzc{d} \delta_L(\mathbf{x})}\right|_{\delta_L^*} \nonumber \\
    =\ & \mathcal{F}'(\delta_{L,R_L}^*)\ \left.\frac{\mathpzc{d} \delta_{L,R_L}}{\mathpzc{d} \delta_L(\mathbf{x})}\right|_{\delta_L^*} \nonumber \\
    =\ & \mathcal{F}'(\delta_{L,R_L}^*)\ \left( W_{R_L} (\mathbf{x}) + \left.\frac{\dd \delta_{L,R'}^*}{\dd R'}\right|_{R_L} \left.\frac{\mathpzc{d} R_L}{\mathpzc{d} \delta_L(\mathbf{x})}\right|_{\delta_L^*} \right)\nonumber \\
    =\ & \mathcal{F}'(\delta_{L,R_L}^*)\ \left( W_{R_L} (\mathbf{x}) + \left.\frac{\dd \delta_{L,R'}^*}{\dd R'}\right|_{R_L} \frac{R^3}{3R_L^2} \left.\frac{\mathpzc{d} \delta_R}{\mathpzc{d} \delta_L(\mathbf{x})}\right|_{\delta_L^*} \right)
\end{align}
\begin{align}
 \Rightarrow\ \lambda\left.\frac{\mathpzc{d} \delta_R}{\mathpzc{d} \delta_L(\mathbf{x})}\right|_{\delta_L^*} =\ & \lambda\frac{\mathcal{F}'(\delta_{L,R_L}^*)\ W_{R_L} (\mathbf{x})}{1 - \mathcal{F}'(\delta_{L,R_L}^*) \left.\frac{\dd \delta_{L,R'}^*}{\dd R'}\right|_{R_L} \frac{R^3}{3R_L^2}} \nonumber \\
 =:\ & A_\lambda[\delta_{L}^*]\ W_{R_L} (\mathbf{x})\\
 \label{eq:solution_for_J_star}
 \Rightarrow\ iJ_L^*(\mathbf{x}) =\ & A_\lambda[\delta_{L}^*]\ W_{R_L} (\mathbf{x})\ .
\end{align}
Inserting \eqnref{solution_for_J_star} into \ref{eq:minimisation_B} we also get a simplified expression for $\delta_L^*$ as
\begin{align}
    & \delta_L^*(\mathbf{x}) = \left.\frac{\mathpzc{d} \Phi}{\mathpzc{d} J_L(\mathbf{x})}\right|_{iJ_L^*} \nonumber \\
    =\ & \left.\sum_{n=2}^\infty\ \frac{1}{(n-1)!} \int \prod_{i=1}^{n-1}\ \dd^3 x_i\ J_L(\mathbf{x}_i)\ \xi_{L,n}(\mathbf{x}_1,\ \dots\ , \mathbf{x}_{n-1} , \mathbf{x})\right|_{iJ_L^*}\nonumber \\
    =\ & \sum_{n=2}^\infty\ \frac{A_\lambda^{n-1}}{(n-1)!}\ \langle \delta_{L,R_L}^{n-1}\ \delta_L(\mathbf{x}) \rangle_c
\end{align}
From this we can further deduce the useful relations
\begin{align}
    \label{eq:solution_for_delta_LR_star}
    \delta_{L,R'}^* =\ & \sum_{n=2}^\infty\ \frac{A_\lambda^{n-1}}{(n-1)!}\ \langle \delta_{L,R_L}^{n-1}\ \delta_{L,R'} \rangle_c\\
    \Rightarrow \delta_{L,R_L}^* =\ & \sum_{n=2}^\infty\ \frac{A_\lambda^{n-1}}{(n-1)!}\ \langle \delta_{L,R_L}^{n}\rangle_c\nonumber \\
    \label{eq:solution_for_delta_LRL_star}
    =\ & \left. \frac{\dd \varphi_{L, R_L}(j)}{\dd j}\right|_{j=A_\lambda}& \\
    \label{eq:solution_for_ddelta_LR_dR_star}
    \Rightarrow \left.\frac{\dd\delta_{L,R'}^*}{\dd R'}\right|_{R_L} =\ & \sum_{n=2}^\infty\ \frac{A_\lambda^{n-1}}{(n-1)!}\ \left\langle \delta_{L,R_L}^{n-1}\ \left.\frac{\dd\delta_{L,R'}}{\dd R'}\right|_{R_L} \right\rangle_c\nonumber\\
    =\ & \frac{1}{A_\lambda} \left. \frac{\dd \varphi_{L, R'}(j)}{\dd R'} \right|_{R'=R_L,\ j=A_\lambda}\ ,
\end{align}
where we have defined the cumulant generating function of the linear, spherically averaged density contrast $\delta_{L,R'}$ as $\varphi_{L, R'}$.

Equations \ref{eq:solution_for_J_star}, \ref{eq:solution_for_delta_LRL_star} and \ref{eq:solution_for_ddelta_LR_dR_star} allow us to obtain $S_\lambda[\delta_L^*, J_L^*]$ by solving a simple 2-dimensional optimization problem. To see this, we first use \ref{eq:solution_for_J_star} in \ref{eq:def_of_action} to get
\begin{align}
    S_\lambda[\delta_L^*, J_L^*] =\ & -\lambda \mathcal{F}(\delta_{L,R_L}^*) + A_\lambda\ \delta_{L,R_L}^* - \varphi_{L, R_L}(A_\lambda)\ .
\end{align}
This is in fact equal to the minimum of the 2-dimensional function
\begin{equation}
\label{eq:2D_function_to_be_minimized}
    s_\lambda(\delta, j) = -\lambda \mathcal{F}(\delta) + j \delta - \varphi_{L, R(1+\mathcal{F}(\delta))^{1/3}}(j)\ ,
\end{equation}
since solving
\begin{equation}
    \left.\frac{\partial s_\lambda}{\partial \delta}\right|_{\delta^*, j^*} =\ \ 0\ \ = \left.\frac{\partial s_\lambda}{\partial j}\right|_{\delta^*, j^*}
\end{equation}
leads to
\begin{align}
    j^* =\ & \lambda \mathcal{F}'(\delta^*) + \left.\frac{\dd \varphi_{R'}(j^*)}{\dd R'}\right|_{R'=R_L}\ \frac{R^3}{3R_L^2}\ \mathcal{F}'(\delta^*)\\
    \delta^* =\ & \left. \frac{\dd \varphi_{R(1+\mathcal{F}(\delta))^{1/3}}(j)}{\dd j}\right|_{j=j^*}
\end{align}
which - as can be seen from \ref{eq:solution_for_delta_LRL_star} and \ref{eq:solution_for_ddelta_LR_dR_star} - has the solutions 
\begin{align}
    \delta^* = \delta_{L,R_L}^*\ ,\ j^* = A_\lambda\ .
\end{align}

We have shown that a spherically symmetric ansatz for the configurations $\delta_L^*$ and $J_L^*$ can extremize the action $S_\lambda$ and that the corresponding extreme values $S_\lambda[\delta_L^* , J_L^*]$ can be obtained by a simple 2-dimensional optimisation problem. We have not yet shown that these $\delta_L^*$ and $J_L^*$ are indeed global minima of the action. We however note that \citet{Valageas2002} has shown that for Gaussian initial conditions there exists a range of values of $\lambda$ for which the spherically symmetric saddle point is a global minimum. Especially, the Hessian of the action $S_\lambda$ is positive definite for these values of $\lambda$. Small deviations from Gaussian initial conditions will change the location of this saddle point, but because of continuity there will be a range of deviations for which the Hessian at this point is still positive definite, such that it is still a minimum of the action. In fact, because of continuity this will also stay a global minimum for sufficiently small deviations from Gaussian initial conditions.

Note that for large enough $\lambda$ the action $S_\lambda[\delta_L, J_L]$ can indeed have two extrema - even for Gaussian initial conditions \citep{Valageas2002, Bernardeau2014}. This leads to a 2-branch structure in the cumulant generating function where the 2nd branch governs the extreme rare-events tail of the PDF. This is an additional reason for why we avoid the high-density tail here.

\subsection{Rescaling to the nonlinear variance}
\label{sec:non_linear_variance_rescaling}

 It has been shown \citep{Fosalba1998, Valageas2002, Valageas2002V} that the saddle-point approximation gives the cumulants of the non-linear density contrast at leading order in perturbation theory. This especially means that the above approximation for the CGF yields the wrong variance of density fluctuations even in the mildly non-linear regime. It is hence much better to consider the re-scaled cumulant generating function \citep{BernardeauReimberg2016}
\begin{equation}
\label{eq:definition_of_rescaled_CGF}
    \tilde{\varphi}_R(\lambda)\ \equiv\ \sigma_{R}^2\ \varphi_R\left(\frac{\lambda}{\sigma_{R}^2}\right)\ =\ \sum_n\ S_n\ \frac{\lambda^n}{n!}\ ,
\end{equation}
where $\sigma_{R}^2$ is the variance of density fluctuations and 
\begin{equation}
\label{eq:defSn}
    S_n\ \equiv\ \frac{\langle \delta_R^n \rangle_c}{\sigma_{R}^{2(n-1)}}\ .
\end{equation}
The reduced cumulants (or hierarchical coefficients) $S_n$ are significantly less sensitive to non-linear evolution than the raw cumulants \citep{BernardeauReview, Bernardeau2014, Bernardeau2015}. This is because tidal terms that are not captured by the leading order term in \eqnref{saddle_point_approximation_for_CGF} are largely erased by smoothing effects in the reduced cumulants, both for Gaussian and non-Gaussian initial conditions \citep{Fosalba1998,Gaztanaga1998}.

Hence, our modelling strategy in this paper is to compute the rescaled CGF at leading order in perturbation theory by calculating the first term in \eqnref{saddle_point_approximation_for_CGF} and using the linear variance $\sigma_{L,R}^2$ in \eqnrefnospace{definition_of_rescaled_CGF} to compute the leading order, re-scaled cumulant generating function $\tilde\varphi_R^{\rm l.o.}(\lambda)$. The final non-linear CGF $\phi_R(\lambda)$ is then obtained by inverting \eqnref{definition_of_rescaled_CGF} with $\sigma_{R}^2$ as a free parameter, \ie
\begin{equation}
    \phi_R(\lambda) = \frac{1}{\sigma_{R}^2}\ \tilde\varphi_R^{\rm l.o.}(\sigma_{R}^2 \lambda)\ .
\end{equation}

As can be seen in \figref{PDF}, this procedure reproduces not only the variance but the overall shape of the density PDF observed in N-body simulations. This re-scaling to the non-linear variance has also been successfully applied to large scale structure data \citep{Friedrich2018, Gruen2018}. Recent findings of \eg\citet{Foreman2019} suggest that even effects of baryonic physics might propagate to higher order moments of the density field primarily through their impact on 2-point statistics. Nevertheless, this re-scaling must be considered a weak point of our modelling, since it has no justification from first principles. Fortunately, \citet{Ivanov2019} have shown that the next-to-leading order term in \eqnref{saddle_point_approximation_for_CGF} can be calculated explicitly, which may eliminate the need to re-scale the CGF by the ratio of linear to non-linear variance (or at least: make the re-scaling even more accurate). 

Note that the non-linear variance itself is also affected by primordial non-Gaussianity through mode-coupling. In particular, it generates an additional odd-order term in the 1-loop power spectrum, $$P(k,z)=D^2(z)P_{\rm L}(k)+P^{(12)}(k,z)+[P^{(22)}+P^{(13)}](k,z)\,,$$ which is absent for Gaussian initial conditions and reads \citep{Taruya2008}
\begin{equation}
    \label{eq:PNLcorr}
    P^{(12)}(k,z)=2 D^3(z) \int\!\!\!\! \frac{\text{d}^3q}{(2\pi)^3} F^{(2)}_{\rm sym}(q,k-q)B_0(-k,q,k-q)\,,
\end{equation} 
where $F^{(2)}_{\rm sym}$ is the symmetrized kernel for the second order perturbative solution of the fluid equations \citep[see e.g. equation~(45) in][]{BernardeauReview}. By marginalising over the non-linear variance we are hence ignoring part of the information about primordial non-Gaussianity that is contained in the density PDF. At the smoothing scales considered here this effect is largest for the orthogonal bispectrum template.

\subsection{Application: primordial bispectrum shapes}
\label{sec:bispec}

If the initial density field was not drawn from a Gaussian distribution, then the linear density contrast today will have a non-zero 3-point function 
\begin{equation}
    \xi_{3,L}(\mathbf{x}_1 ,\mathbf{x}_2 ,\mathbf{x}_3) = \langle \delta_L(\mathbf{x}_1) \delta_L(\mathbf{x}_2) \delta_L(\mathbf{x}_3) \rangle_c\ .
\end{equation}
The Bispectrum $B_{L}(k_1 ,k_2 ,k_3)$ is defined through the Fourier transform of $\xi_{3,L}$ in each of its arguments (see \apprefnospace{Fourier_conventions} for our Fourier conventions) as
\begin{align}
&\langle \tilde\delta_L(\mathbf{k}_1) \tilde\delta_L(\mathbf{k}_2) \tilde\delta_L(\mathbf{k}_3)\rangle_c\nonumber \\
 \equiv\ & (2\pi)^3 \delta_D(\mathbf{k}_1+\mathbf{k}_2 + \mathbf{k}_3)\ B_L(k_1, k_2, k_3)\ .
\end{align}
The skewness of the linear density field averaged over a spherical top-hat filter of radius $R$ is then given in either real space or Fourier space by
\begin{align}
    \langle \delta_{L,R}^3 \rangle_{c} =\ & \left\langle\prod_{i=1}^3 \int \dd^3 x_i\ W_R(\mathbf{x}_i) \delta_L(\mathbf{x}_i)\right\rangle_c \nonumber \\
     =\ & \left\langle\prod_{i=1}^3 \int \frac{\dd^3 k_i}{(2\pi)^3}\ \tilde W_R(\mathbf{k}_i) \tilde\delta_L(\mathbf{k}_i)\right\rangle_c \ ,
\end{align}
where $W_R(\mathbf{x})$ and $\tilde W_R(\mathbf{x})$ are given in \eqnref{defsmooth} and \eqnref{filter} respectively. The Fourier space expression can be simplified to
\begin{align}
& \langle \delta_{L,R}^3 \rangle_{c} =  \left\langle\prod_{i=1}^3 \int \frac{\dd^3 k_i}{(2\pi)^3}\ \tilde W_R(\mathbf{k}_i) \tilde\delta_L(\mathbf{k}_i)\right\rangle_c \nonumber \\
  &\ \ \ \ \   = \int \frac{\dd^3 k_1\dd^3 k_2}{(2\pi)^6}\ \tilde W_R(k_1) \tilde W_R(k_2) \tilde W_R(k_3) \ B_L(k_1, k_2, k_3)\ ,
\end{align}
where in the last line we set
\begin{align}
    k_3 \equiv |\mathbf{k}_1+\mathbf{k}_2| = \sqrt{k_1^2 + k_2^2 + 2\mu k_1 k_2}
\end{align}
and $\mathbf{k}_1\cdot\mathbf{k}_2 \equiv \mu k_1 k_2$ . The above integrand only depends on the angle between the two remaining wave vectors. Hence we can perform the angular integral for one of these to get
\begin{align}
\label{eq:skewness_in_terms_of_B_L}
    & \langle \delta_{L,R}^3 \rangle_{c}\nonumber \\
    =\ & \frac{2}{(2\pi)^4}\int \dd \ln k_1\ k_1^2\ \tilde W_R(k_1)\ \int \dd \ln k_2\ k_2^2\ \tilde W_R(k_2)\ \times\nonumber \\
    & \times\ \int_{|k_1 - k_2|}^{(k_1 + k_2)} \dd \ln k_3\ k_3^2\ \tilde W_R(k_3)\ B_L(k_1, k_2, k_3)\ .\nonumber \\
\end{align}
We would now like to express $B_L(k_1, k_2, k_3)$ in terms of a Bispectrum of primordial potential fluctuations, $B_\phi(k_1, k_2, k_3)$. The initial potential fluctuations $\tilde\phi_i(\mathbf{k})$ are related to the linear density contrast today via the Poisson equation
\begin{align}
   \tilde\delta_L(\mathbf{k}) = \frac{-2\tilde T(k)k^2}{3\Omega_m H_0^2}\ \frac{D(z_i)}{a(z_i)}\ \tilde\phi_i(\mathbf{k})\ ,
\end{align}
where $\tilde T(k)$ is the transfer function (defined such that $\tilde T(k\rightarrow 0) = 1$), $D(z)$ and $a(z)$ are the linear growth factor and scale factor (with $D = 1 = a$ today) and $z_i$ is some redshift during matter domination. Now common templates for $B_\phi$ contain terms of the form
\begin{equation}
\label{eq:terms_in_primordial_bispectrum}
    B_\phi(k_1, k_2, k_3) \supset P_\phi(k_1)^{\alpha_1} P_\phi(k_2)^{\alpha_2} P_\phi(k_3)^{\alpha_3}
\end{equation}
such that $\alpha_1+\alpha_2+\alpha_3 = 2$ \citep{Fergusson2009, Scoccimarro2012, Uhlemann2018c}. In terms of the linear matter power spectrum today these terms read
\begin{align}
    & P_\phi(k_1)^{\alpha_1} P_\phi(k_2)^{\alpha_2} P_\phi(k_3)^{\alpha_3} \nonumber \\
    =\ & \left(\frac{3\Omega_m H_0^2}{2} \frac{a(z_i)}{D(z_i)}\right)^4 P_L(k_1)^{\alpha_1} P_L(k_2)^{\alpha_2} P_L(k_3)^{\alpha_3}\ \times \nonumber \\
    & \times\ \tilde T(k_1)^{-2\alpha_1} \tilde T(k_2)^{-2\alpha_2} \tilde T(k_3)^{-2\alpha_3}\ k_1^{-4\alpha_1} k_2^{-4\alpha_2} k_3^{-4\alpha_3}\ .
\end{align}
Hence, the linear density Bispectrum will contain terms like
\begin{align}
    B_L(k_1, k_2, k_3) \supset\ & \frac{-3\Omega_m H_0^2a(z_i)}{2D(z_i)}\ P_L(k_1)^{\alpha_1} P_L(k_2)^{\alpha_2} P_L(k_3)^{\alpha_3} \nonumber \\
    & \times\ \tilde T(k_1)^{1-2\alpha_1} \tilde T(k_2)^{1-2\alpha_2} \tilde T(k_3)^{1-2\alpha_3}\ \times \nonumber \\
    & \times\ k_1^{2-4\alpha_1} k_2^{2-4\alpha_2} k_3^{2-4\alpha_3}\ .
\end{align}
Inserting this into our above equations for the skewness we arrive at
\begin{align}
\label{eq:skewness_terms_with_P_delta}
    & \langle \delta_{L,R}^3 \rangle_{c} \supset \frac{-3\Omega_m H_0^2}{(2\pi)^4} \frac{a(z_i)}{D(z_i)}\ \times \nonumber \\
    & \times\ \int \dd \ln k_1\ k_1^{4-4\alpha_1}\ \tilde T(k_1)^{1-2\alpha_1}\ \tilde W_R(k_1)\ P_L(k_1)^{\alpha_1}\ \times\nonumber \\
    & \times\ \int \dd \ln k_2\ k_2^{4-4\alpha_2}\ \tilde T(k_2)^{1-2\alpha_2}\ \tilde W_R(k_2)\ P_L(k_2)^{\alpha_2}\ \times\nonumber \\
    & \times\ \int_{\min(k_1, k_2)}^{\max(k_1, k_2)} \dd \ln k_3\ k_3^{4-4\alpha_3}\ T(k_3)^{1-2\alpha_3}\ \tilde W_R(k_3)\ P_L(k_3)^{\alpha_3}\ .\nonumber \\
\end{align}
In order to calculate the skewness of the linear density field as a function of $R$ we now only have to specify what terms of the form of (\ref{eq:terms_in_primordial_bispectrum}) are present in the primordial bispectrum. Following \eg \citet{Uhlemann2018c} (see also the other references given in this section) we consider local, orthogonal and equilateral shapes for the bispectrum which are given by 
\begin{align}
\label{eq:local_template}
    B_\phi^{\mathrm{loc}}(k_1, k_2, k_3)  =\ & -2f_{\mathrm{NL}}^{\mathrm{loc}} (P_\phi(k_1)P_\phi(k_2) + 2\ \mathrm{permutations}) \\
\label{eq:equi_template}
    B_\phi^{\mathrm{equi}}(k_1, k_2, k_3)  =\ & 6f_{\mathrm{NL}}^{\mathrm{equi}}\ \left[\ (P_\phi(k_1)P_\phi(k_2) + 2\ \mathrm{perm.}) \right. \nonumber \\
    & + 2 P_\phi^{2/3}(k_1)P_\phi^{2/3}(k_2)P_\phi^{2/3}(k_3) \nonumber \\
    &\left. - (P_\phi(k_1)P_\phi^{2/3}(k_2)P_\phi^{1/3}(k_3)  + 5\ \mathrm{perm.})\right]\nonumber \\
    \\
\label{eq:ortho_template}
    B_\phi^{\mathrm{ortho}}(k_1, k_2, k_3)  =\ & 6f_{\mathrm{NL}}^{\mathrm{ortho}}\ \left[\ 3 (P_\phi(k_1)P_\phi(k_2) + 2\ \mathrm{perm.}) \right. \nonumber \\
    & + 8 P_\phi^{2/3}(k_1)P_\phi^{2/3}(k_2)P_\phi^{2/3}(k_3) \nonumber \\
    &\left. - 3(P_\phi(k_1)P_\phi^{2/3}(k_2)P_\phi^{1/3}(k_3)  + 5\ \mathrm{perm.})\right]\ .\nonumber \\
\end{align}
Note that these expression differ in sign from equations given in other publications \citep[\eg][]{Mao2014, Uhlemann2018b}. This is because here $\phi$ denotes the Newtonian potential and not Bardeen's potential \citep{Salopek1990}.


\subsection{Summary of our recipe for the matter PDF}
\label{sec:recipe}

To summarize, the model predictions in Figures~\ref{fi:PDF}-\ref{fi:moment_residuals_summary} and the forecasts presented in Figures~\ref{fi:constraints_joint}-\ref{fi:constraints_R15} are calculated as follows:
\begin{enumerate}
    \item computing skewness of the linear density contrast $\langle \delta_{L,R}^3 \rangle_c$ from \eqnref{skewness_in_terms_of_B_L}, using either of the primordial bispectrum templates given in Equations \ref{eq:local_template}-\ref{eq:ortho_template}
    \item approximating the cumulant generating function of the linear density field in terms of the linear variance and skewness as
    \begin{equation}
    \label{eq:approx_var_L_as_cubic}
        \varphi_{L,R}(\lambda) \approx \langle \delta_{L,R}^2 \rangle_c \frac{\lambda^2}{2!} + \langle \delta_{L,R}^3 \rangle_c \frac{\lambda^3}{3!}
    \end{equation}
    \item using this in \eqnref{2D_function_to_be_minimized} to calculate the late-time cumulant generating function (applying also the variance re-scaling described in \secrefnospace{non_linear_variance_rescaling})
    \item numerically evaluating the inverse Laplace transform of \eqnref{PDF_as_Laplace_transform} to obtain the late-time density PDF.
\end{enumerate}

\subsection{Numerical calculation of the CGF and extraction of individual cumulants}
\label{sec:numerical_details_moment_extraction}

\begin{centering}
\begin{figure}
  \includegraphics[width=0.47\textwidth]{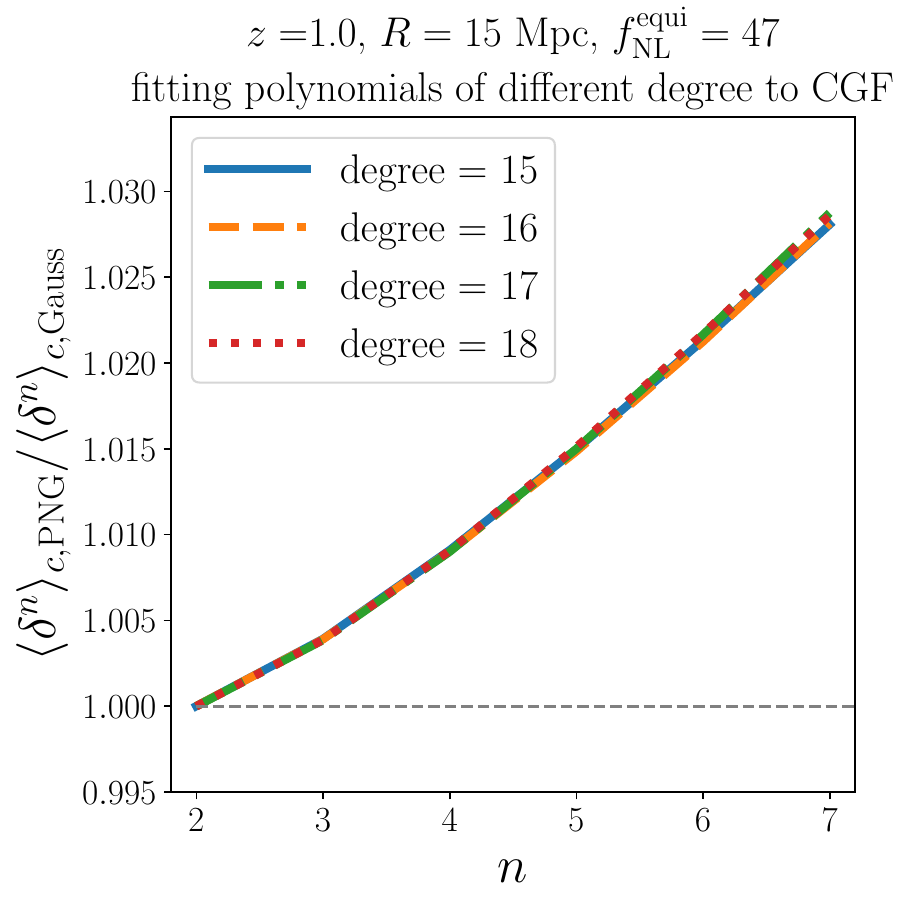}
   \caption{Theoretical predictions for the ratio of the $n$th connected moments of the late-time density field from non-Gaussian initial conditions to Gaussian initial conditions. The individual moments have been obtained from our numerically calculated cumulant generating functions by the polynomial fitting procedure described in \secrefnospace{numerical_details_moment_extraction}. We again assume $f_{\mathrm{NL}}^{\mathrm{equi}}=47$, which corresponds to the $1\sigma$ uncertainty of \citetalias{Planck2018PNG}.}
  \label{fi:moment_residuals_from_polynomial_fit}
\end{figure}
\end{centering}

In \secref{minimizing_the_action} we demonstrated that the leading order approximation of the cumulant generating function 
\begin{equation}
\label{eq:varphi_R_definition_numerical_section}
    \varphi_R(\lambda) = \sum_{n=2}^\infty \langle \delta_R^n\rangle_c\ \frac{\lambda^n}{n!}
\end{equation}
is given by the minimum of the 2-dimensional function
\begin{equation}
\label{eq:2D_function_to_be_minimized_v2}
    s_\lambda(\delta, j) = -\lambda \mathcal{F}(\delta) + j \delta - \varphi_{L, R(1+\mathcal{F}(\delta))^{1/3}}(j)\ .
\end{equation}
Minimizing this function amounts to solving the equations
\begin{align}
    \label{eq:numerical_tricks_1}
    j^* =\ & \lambda \mathcal{F}^{\ '}(\delta^*) + \left.\frac{\dd \varphi_{R'}(j^*)}{\dd R'}\right|_{R'=R_L}\ \frac{R^3}{3R_L^2}\ \mathcal{F}^{\ '}(\delta^*)\\
    \label{eq:numerical_tricks_2}
    \delta^* =\ & \left. \frac{\dd \varphi_{L,R(1+\mathcal{F}(\delta^*))^{1/3}}(j)}{\dd j}\right|_{j=j^*}\ .
\end{align}
However, this is not what we do in practice. Instead, it is much easier to proceed as follows:
\begin{enumerate}
    \item For an array of values for $\delta^*$, calculate the corresponding arrays $\mathcal{F}(\delta^*)$ and $\mathcal{F}^{\ '}(\delta^*)$ using \apprefnospace{spherical_collapse}. Also compute the variance and skewness of the linear density field at the radii $R_L(\delta^*) = R(1+\mathcal{F}(\delta^*))^{1/3}$ .
    \item For each value of $\delta^*$ invert \eqnref{numerical_tricks_2} to obtain $j^* = j^*(\delta^*)$. Since we approximate $\varphi_{L,R}(\lambda)$ as a cubic function in \eqnref{approx_var_L_as_cubic}, this just corresponds to solving a quadratic equation.
    \item Now invert \eqnref{numerical_tricks_1} to obtain $\lambda = \lambda(\delta^*)$ .
    \item Finally, obtain the CGF from $\varphi_R(\lambda) = s_\lambda(\delta^*, j^*)$ .
\end{enumerate}
This is a constructive procedure to obtain the cumulant generating function which does not actually require one to solve any optimization problem.

The above steps yield the cumulant generating function $\varphi_R(\lambda)$ on a grid of values of $\lambda$. In order to extract individual cumulants from that (\ie the Taylor coefficients in \eqnrefnospace{varphi_R_definition_numerical_section}) one might attempt to simply fit a polynomial of finite degree in $\lambda$ to the CGF $\varphi_R(\lambda)$. This is however highly unstable, mainly because positive values of $\delta^*$ are mapped to a very small interval of $\lambda(\delta^*)$ compared to the interval that negative values of $\delta^*$ are mapped to \citep[see \eg Figure 3 of][]{Valageas2002}. A more robust way to extract individual moments is to first define the new variable \citep{Bernardeau2000, Valageas2002, Bernardeau2015}
\begin{equation}
    \tau(\delta^*) = \frac{\delta^*}{\langle (\delta_{L,R_L(\delta^*)})^2 \rangle_c}\ .
\end{equation}
Then one proceeds as follows:
\begin{enumerate}
    \item Fit a polynomial of finite order $N$ in $\tau$ to both $\lambda(\tau)$ and  $\varphi_R(\lambda(\tau))$. 
    \item Determine the $N\times N$ Bell-Jabotinsky matrices \citep{Jabotinsky1963} of $\lambda(\tau)$ and  $\varphi_R(\lambda(\tau))$ \wrt $\tau$
    \begin{equation}
        \left(B^{\lambda | \tau}\right)_{k,\ell} = \frac{1}{k!} \left. \frac{\dd^k \lambda^\ell}{\dd \tau^k} \right|_{\tau = 0}\ ;\ \left(B^{\phi | \tau}\right)_{k,\ell} = \frac{1}{k!} \left. \frac{\dd^k \phi^\ell}{\dd \tau^k} \right|_{\tau = 0}\ .
    \end{equation}
    \item The Bell-Jabotinsky matrix of $\phi$ \wrt $\lambda$ (whose column $\ell=1$ contains the cumulants of order $k$ divided by $k!$) is then given by
    \begin{equation}
        \mathbf{B}^{\phi | \lambda} = \mathbf{B}^{\phi | \tau} \cdot \left(\mathbf{B}^{\lambda | \tau} \right)^{-1}\ .
    \end{equation}
\end{enumerate}

The cumulant ratios shown in \figref{moment_residuals_summary} and \figref{moment_residuals_from_polynomial_fit} were computed along these lines. \Figref{moment_residuals_from_polynomial_fit} shows that this procedure converges up to the 7th cumulant for a polynomial degree of about $N=16$. However, this statement is highly dependent on the redshift and smoothing scale at which the density field is considered and the above steps should not be carried out as a black box! Instead, we recommend comparisons such as \Figref{moment_residuals_from_polynomial_fit} as tests for robustness.

\section{Simulated data and covariance matrix}
\label{sec:covariance}

\subsection{Data from N-body simulations}
\label{sec:sims}
In the following paragraphs we summarise the different N-body simulations used in our work. Their cosmological parameters and available primordial non-Gaussianty model amplitudes are compared in Table~\ref{tab:cosmoparam}.

\subsubsection{Quijote simulations}

The Quijote simulations \citep[described in][]{Navarro2019} are a large suite of N-body simulations developed for quantifying the cosmological information content of large scale structure observables. They contain 43100 simulations spanning over 7000 cosmological models with variation in $\nu\Lambda$CDM parameters. At a single redshift, the combined number of particles in the simulation suite is > 8.5 Trillion, with a combined volume of 43100 $({\rm Gpc}/h)^3$. The simulations follow the gravitational evolution of $N^3$ particles ($2\times N^3$ for simulations that include massive neutrinos) over a co-moving volume of 1 $({\rm Gpc}/h)^3$ starting from $z=127$. Here $N$ takes the values: $N=256$ (low-resolution), $N=512$ (fiducial-resolution) and $N=1024$ (high-resolution). Overall, 15000 simulations are provided for the fiducial resolution, allowing to accurately estimate the covariance matrices of cosmological summary statistics. We refer the reader to \cite{Navarro2019} for further details.
In this paper we investigate to the probability distribution function (PDF) of the cosmic matter density field. The PDFs have been computed from the Quijote simulations as follows. First, particle positions and masses are assigned to a regular grid with $512^3$ cells using the Cloud-in-Cell (CIC) mass assignment scheme ($1024^3$ for the high-resolution simulations). Next, the value of the overdensity field in each grid cell is computed by dividing the mass of each cell by the average mass. Finally, the PDF is estimated by calculating the fraction of cells that lie in a given overdensity bin, over the width of the overdensity bin itself.

The covariance matrix used for our forecasts is entirely obtained from the Quijote simulations. We combine simulated measurements of the PDF and the cumulants at smoothing scales $R=15,\ 30$Mpc$/h$ into data vectors $\mathbf{d}_i\ (i = 1,\ \dots\ , 15000)$ to estimate the covariance as
\begin{equation}
    \hat C = \frac{1}{N_{\mathrm{sim}}-1} \sum_i (\mathbf{d}_i - \bar{\mathbf{d}}) (\mathbf{d}_i - \bar{\mathbf{d}})^T\ ,
\end{equation}
where $\bar{\mathbf{d}}$ is the mean of all measurements and $N_{\mathrm{sim}}=15000$ for Quijote. In \figref{correlation_matrix} we show the corresponding correlation matrix of measured PDF histograms as well as of the measured variance, skewness and kurtosis. It can be seen there, that at both low and high densities the PDFs are largely anti-correlated with the height of the PDF peak. Related to that, also measurements of the variance are anti-correlated with the peak height of the density PDF. In contrast, all cumulants are positively correlated to the probabilities of low and high densities in the PDFs. However, the onset of the positive correlation moves to higher densities when going to higher cumulant orders.

\subsubsection{Simulations by Nishimichi et al. for local primordial non-Gaussianity}

For \figref{PDF_residuals_f_NL_100} we also investigated a set of simulations that have previously been studied by \citet{Uhlemann2018b}. These simulations are based in a non-Gaussian initial condition generator developed by \citet{Nishimichi2012} as well as a parallel code developed by \citet{Valageas2011}. The simulations contain $2048^3$ particles in a box of length $4096$ Mpc$/h$. One such box has been run with Gaussian initial conditions and two boxes have been run with a local-type primordial bispectrum of amplitude $f_{\rm NL}=\pm 100$ respectively. 
Such exaggerated amplitudes are ruled out by \citetalias{Planck2018PNG}, but we anyway only use these simulations as a test for our PDF modelling approach. The cosmological parameters of the simulations are summarized in Table~\ref{tab:cosmoparam}. We use a snapshot of these simulations at $z=1$. The density field in the simulations is evolved using the Tree-PM code Gadget2 \citep{Springel2005,Springel2001}. The results shown here measure the PDF based on a clouds-in-cells (CIC) mass assignment with $1280^3$ grid points that includes window and aliasing corrections.

\begin{table}
\begin{tabular}{c|cccc|c}
simulation & $\Omega_m$ & $h$ & $n_s$ & $\sigma_8$ & $f_{\rm NL}$\\\hline
Quijote & 0.3175 & 0.6811 & 0.9624 & 0.834 & - \\ \hline
Nishimichi & 0.279 & 0.701 & 0.96 & 0.8157 & $\{\pm 100,0\}$ \\
Oriana & 0.25 & 0.7 & 1.0 & 0.8 & $\{100,-400\}$\\
\end{tabular}
\caption{Cosmological parameters of the considered simulations, all run for a flat $\Lambda$CDM cosmology with $\Omega_\Lambda=1-\Omega_m$. The values quoted for the amplitude of primordial non-Gaussianity $f_{\rm NL}$ are $\{f_{\rm NL}^{\rm loc},f_{\rm NL}^{\rm equi/ortho}\}$.}
\label{tab:cosmoparam}
\end{table}

\subsubsection{The Oriana simulations for three different primordial bispectrum shapes}

For our comparison in \figref{PDF_residuals_f_NL_100} we also used simulated data from the large volume ``Oriana'' realisations of the Large Suite of Dark Matter Simulations project \citep[LasDamas;][]{McBride2009}. The simulation was run for cosmological parameters similar to WMAP year 5 \citep{Komatsu2009}, as summarised in Table~\ref{tab:cosmoparam}. The Oriana simulations evolve $1280^3$ dark matter particles in a cubic volume of $(2.4Gpc/h)^3$, resulting in a particle mass of $45.7 \times 10^{10} M_{\odot}/h$. The simulation seeds are generated from second-order Lagrangian perturbation theory (2LPT) initial conditions \citep{Scoccimarro1998,Crocce2006} and evolved from a starting redshift of $z_{init} = 49$ to $z=0$ using the Gadget-2 code \citep{Springel2005}, with a gravitational force softening of $53$ kpc$/h$.  

Of the Oriana suite we analyse here one box with Gaussian initial conditions and three realisations initialised with primordial non-Gaussianity models of either local ($f_{\rm NL}^{\rm loc}=100$), equilateral ($f_{\rm NL}^{\rm equi}=-400$) or orthogonal ($f_{\rm NL}^{\rm orth}=-400$) initial bispectra \citep{Scoccimarro2012}.
Such exaggerated amplitudes are ruled out by \citetalias{Planck2018PNG}, but we anyway only test our PDF modelling approach with these simulations.

\subsection{Log-normal simulations}
\label{sec:lognormal_sims}

We also explore a cheap way of estimating the covariance matrix (for the purpose of future data analyses) by generating zero-mean shifted log-normal density fields \citep{Hilbert2011, Xavier2016}. 

As described \eg in \citet{Hilbert2011, Xavier2016} we consider the density contrast $\delta(\mathbf{x})$ to be a \emph{zeros-mean shifted log-normal random field}, which is given in terms of a Gaussian random field $g(\mathbf{x})$ as
\begin{equation}
\label{eq:lognormal_definition}
    \delta(\mathbf{x}) = \delta_0\left[e^{g(\mathbf{x})} - 1 \right] \ ,
\end{equation}
where $\delta_0$ is a free parameter that can be used to tune certain higher-order statistical properties of $\delta$. Demanding that $\langle \delta \rangle = 0$ fixes the mean value $\mu_g$ and variance $\sigma_g^2$ of the Gaussian field to obey the relation
\begin{equation}
    \mu_g = - \frac{\sigma_g^2}{2}\ .
\end{equation}
The 2-point correlation functions of $\delta(\mathbf{x})$ and $g(\mathbf{x})$ are then related through \citep{Hilbert2011}
\begin{equation}
    \xi_g(x) = \ln \left( 1 + \frac{\xi_\delta(x)}{\delta_0^2} \right)
\end{equation}
For a given choice of $\delta_0 $ and a desired power spectrum of $\delta$, this enables us to calculate the corresponding power spectrum of $g$ and hence to draw it from the appropriate Gaussian distribution.

We follow \citet{Friedrich2018} and \citet{Gruen2018} and choose $\delta_0 $ such that the resulting log-normal field $\delta$ has the same skewness as that calculated from our fiducial PDF model on a scale of $R = 15$Mpc$/h$ (and at $z = 1$). In practice, to perform these steps, we make use of the \verb|python| tool \verb|nbodykit|\footnote{\url{https://nbodykit.readthedocs.io/}} and generate the random fields $\delta(\mathbf{x})$ and $g(\mathbf{x})$ on a grid of volume $L^3 = 1 (\mathrm{Gpc}/h)^3$ and with a number of $512^3$ grid points. Note that an important ingredient in our procedure is a theorem by \citet{Szyszkowicz2009}. They show that averaging a log-normal random field of the form given in (\ref{eq:lognormal_definition}) on a scale $R$ yields a random field that in certain limits is also well described by a log-normal random field, and what is more: by a log-normal random field with the same value of $\delta_0$. This allows us to impose our desired value of $\delta_0$ on the grid scale and obtain the same value also on larger smoothing scales.

The lower left corner of \figref{correlation_matrix} shows the correlation matrix of measurements of the PDF and cumulants of the density field obtained from 400 log-normal simulations generated at the cosmology of the Quijote simulations. The overall structure of correlations seems to be well captured with this simplified approach compared to the correlation matrix obtained from Quijote (upper right corner).

\section{Conclusions \& Discussion}
\label{sec:discussion}

\begin{centering}
\begin{figure}
  \includegraphics[width=0.47\textwidth]{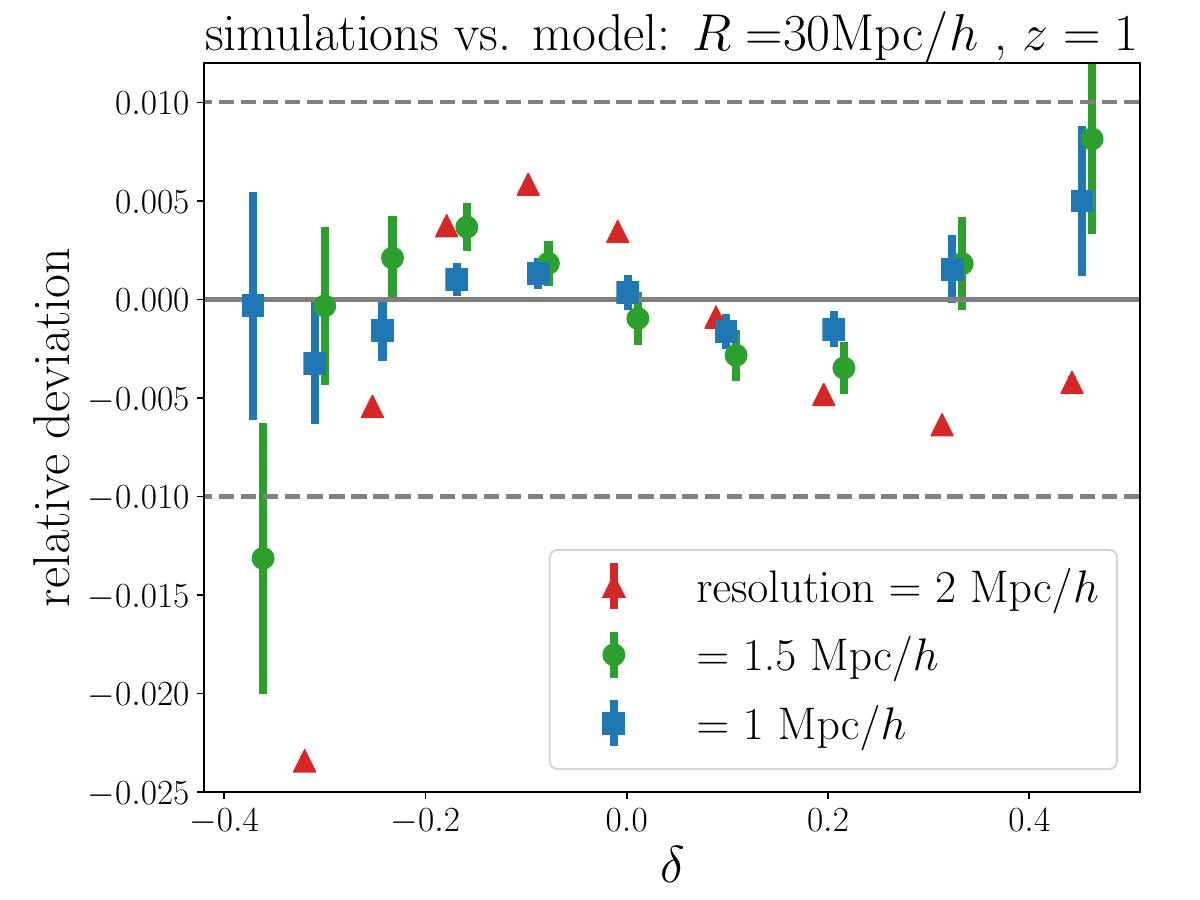}

  \includegraphics[width=0.47\textwidth]{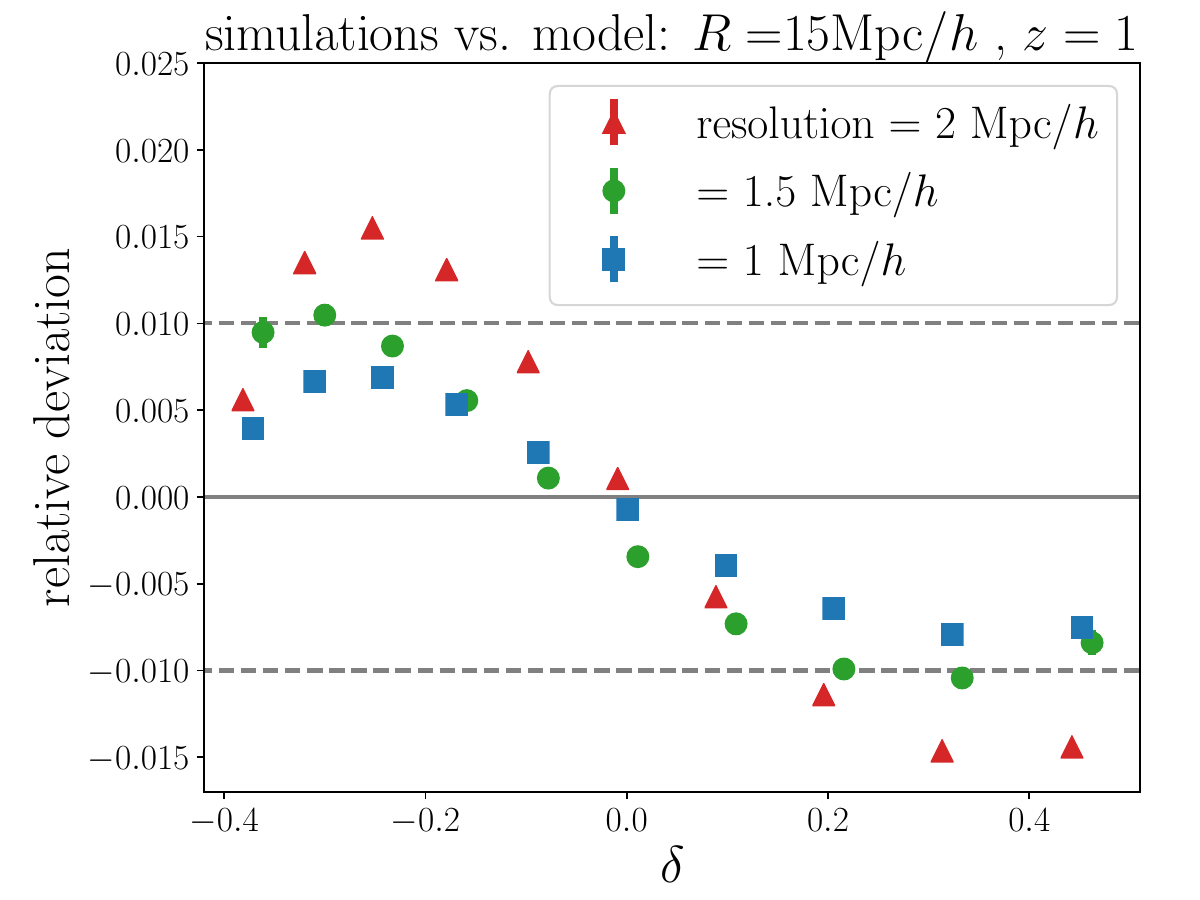}
   \caption{Investigating the impact of simulation resolution (by which we denote the distance of neighboring matter particles in the initial conditions of each simulations) on measurements of the density PDF. Quijote fiducial resolution is shown in red and Quijote high-resolution is shown in blue. We also consider a set of N-body simulations at an intermediate resolution \citep[green points, \cf][]{Baldauf2016}. Note that these simulations where run at a slightly different cosmology than Quijote, which is the reason for the different shapes of the residuals. Error bars denote the errors of the mean of each set of simulations, \ie the actual statistical uncertainties of the points (corresponding to a survey volume of $100(\mathrm{Gpc}/h)^3$ for high-resolution Quijote). \emph{Upper panel:} Relative deviation of the simulations \wrt our analytic prediction of the PDF at $R=30$Mpc$/h$ , $z=1$. \emph{Lower panel:} The same comparison but for $R=15$Mpc$/h$. The main cause of our remaining modelling uncertainty it the inaccuracy of the re-scaling procedure described in \secrefnospace{non_linear_variance_rescaling} (see the discussion in \secrefnospace{discussion}).}  
  \label{fi:Quijote_residuals}
\end{figure}
\end{centering}

In this study we have have quantified the impact of primordial non-Gaussianity on the late-time density PDF and extended existing work \citep{Valageas2002III, Uhlemann2018c} in two ways: First, we presented a new method of modelling the impact of general non-Gaussian initial conditions on the PDF of the late-time density field. This method is an extension of the steepest descent approach by \citet{Valageas2002} for Gaussian initial conditions and requires fewer approximations than existing approaches to model the impact primordial non-Gaussianity on the PDF.

Secondly, we considered the full covariance matrix of measured density PDFs to forecast the statistical power of such measurements in determining the amplitude of different primordial bispectrum shapes. We considered a combined analysis of the PDF on scales of $15$ and $30$Mpc$/h$ at redshift $z=1$ in a survey volume of $V=100$ (Gpc$/h$)$^3$. This is smaller than the effective volume of upcoming surveys such as Spherex with $V_{\mathrm{eff}}\approx 150$ (Gpc$/h$)$^3$ and larger than existing surveys such as BOSS with $V_{\mathrm{eff}}\approx 55$ (Gpc$/h$)$^3$ \citep{Dore2014, Alam2017}. We found that such an analysis can measure the amplitudes of different primordial bispectrum templates with statistical uncertainties of $\Delta f_{\mathrm{NL}}^{\mathrm{loc}} = \pm 3.1$, $\Delta f_{\mathrm{NL}}^{\mathrm{equi}} = \pm 10.0$, $\Delta f_{\mathrm{NL}}^{\mathrm{orth}} = \pm 17.0$, even when treating the non-linear variance of the density field at both smoothing scales as two independent, free parameters. This marginalisation makes our results independent of the amplitude of linear density fluctuations as parametrised by sigma $\sigma_8$. Other cosmological parameters such as $\Omega_m$ have been kept fixed in our analysis \citep[see][for an investigation of the general cosmology dependence of the PDF]{Uhlemann2020}. But we note that \citet{Friedrich2018} and \citet{Gruen2018} have demonstrated how lensing-around-cells can be used to simultaneously obtain information about parameters of a background $\Lambda$CDM model and higher order moments of the density field in a PDF-based analysis. Our work was done in preparation of a combined analysis of the late-time PDF and the early-universe results of \citetalias{Planck2018PNG}. These two cosmological probes have the potential to powerfully complement each other: the CMB can provide information about the background $\Lambda$CDM spacetime, the late-time density PDF contains information about non-linear structure growth and both of them contain independent information about the imprint of primordial non-Gaussianity on the large scale structure.

We want to stress again, that observational data of the large scale structure have already been successfully analysed based on a modelling framework that is closely related to the one presented here \citep{Friedrich2018, Gruen2018}. Still, to harvest the statistical power of the density PDF demonstrated here, a number of problems have to be addressed which we will discuss in the following sections.

\subsection{Precision of our model}
\label{sec:model_precision}

An important question is whether the analytic modelling and/or the simulated data presented here are accurate enough to analyse future large scale structure data. In \figref{Quijote_residuals} we show the relative deviation of PDFs measured in N-body simulations \wrt our theoretical model. The red and blue points use measurements from the fiducial Quijote run (red, spacing of initial particle grid $\approx 2$Mpc$/h$) and from a high-resolution version of Quijote (blue, spacing of initial particle grid $\approx 1$Mpc$/h$). The green points use an additional set of simulations at an intermediate resolution ($\approx 1.5$Mpc$/h$, \cf \citealt{Baldauf2016} as well as our \appref{higher_res_test} for details). Our model for the Quijote simulations was computed for a non-linear variance that gives the best fit to the high-resolution data. For the intermediate resolution data we had to fit the non-linear variance separately, since those simulations were run on a slightly different cosmology.

The upper panel of \figref{Quijote_residuals} shows these residuals for a smoothing scale of $R=30$Mpc$/h$ and the lower panel shows them for a smoothing scale of $R=15$Mpc$/h$. At both scales, going to higher resolution in the simulations improves the fit between model and data. At the $R=30$Mpc$/h$ scale, our model is consistent with the high-resolution simulations at about 1 $\sigma$, with $\chi^2 = 9.9$ (expected: $\approx 6 \pm 3.46$). Overall we have 100 high-resolution boxes of size $(1 \mathrm{Gpc}/h)^3$, \ie the error bars on the blue points in \figref{Quijote_residuals} correspond to the survey volume of $V \approx 100 (\mathrm{Gpc}/h)^3$ that we considered throughout this paper. We hence conclude that at $R=30$Mpc$/h$ our model captures the non-linear evolution of the PDF accurately enough for analysis on these kind of volumes.

For $R=15$Mpc$/h$ our model agrees with the high-resolution data within $1\%$. But this is significantly larger than the statistical uncertainties expected in future large scale structure data! We ran a limited number of simulations with even higher resolution (\cf \apprefnospace{higher_res_test}) to test whether this may be due to inaccuracy of the simulations. Within the limited statistical power of this comparison this does not seem to be the case (\cf \figrefnospace{Tobias_residuals}).

On the modelling side, the most critical approximation we made is the assumption that the reduced cumulants
\begin{equation}
    S_n = \frac{\langle \delta_R^n \rangle_c}{\langle \delta_R^2 \rangle_c^{n-1}}
\end{equation}
are well modelled by the leading order term of \eqnref{saddle_point_approximation_for_CGF} (\ie by tree level perturbation theory) even in the regime where the cumulants $\langle \delta_R^n \rangle_c$ themselves have significant next-to-leading order contribution. This assumption allows us to apply the variance re-scaling described in \secref{non_linear_variance_rescaling} and it is justified by the observation that tidal terms which are not captured by the leading order term are largely erased by smoothing effects in the reduced cumulants \citep{Fosalba1998,Gaztanaga1998}. A way to improve the accuracy of our variance re-scaling (and potentially even make it unnecessary) is to take into account the next-to-leading order term in \eqnref{saddle_point_approximation_for_CGF}. \citet{Ivanov2019} have demonstrated how to calculate this term for Gaussian initial conditions.

Finally, \figref{PDF_residuals_f_NL_100} demonstrates that there is also residual disagreement between how the PDF responds to primordial non-Gaussianity in N-body simulations and in our model. We note that the simulations investigated in this figure were only available to us in resolutions similar to the fiducial Quijote run. And \figref{Quijote_residuals} indicates that at our smoothing scales this is insufficient to obtain accurate density PDFs. Also, the simulations considered in \figref{PDF_residuals_f_NL_100} exhibit exaggerated amounts of primordial non-Gaussianity. This may impact the performance of the saddle point approximation derived in \secref{model} as well as the assumption that the primordial CGF is well approximated by a cubic function (\eqnrefnospace{approx_var_L_as_cubic}).

In summary, we conclude the following regarding the precision of our analytic modelling:
\begin{itemize}
    \item At $R=30$Mpc$/h$ and $z=1$ our model is consistent with high-resolution simulated data that corresponds to an overall survey volume of $V \approx 100 (\mathrm{Gpc}/h)^3$.
    \item At $R=15$Mpc$/h$, there is a residual disagreement of $\lesssim 1\%$ between our model and the Quijote high-resolution run. We will investigate this in future work using even higher resolution N-body simulations.
    \item There is also residual disagreement between simulations and our model regarding the response of the PDF to primordial non-Gaussianity (\figrefnospace{PDF_residuals_f_NL_100}). This is likeli because of the limited resolution of the considered simulations as well as their exaggerated values of $f_{\mathrm{NL}}$.
    \item A promising route to improve our analytic modelling is to extend the formalism of \citet{Ivanov2019} for non-Gaussian initial conditions. A major obstacle in applying such an approach in real data analysis is however that it is computationally very expensive.
\end{itemize}

\subsection{Redshift uncertainties}

In this paper we have only considered the 3D density field at one redshift. In real situations we cannot access this kind of information for a number of reasons. Firstly, any cosmological observations take place along our past light-cone. Secondly, peculiar velocities of tracers introduce distortions in the mapping between observed redshift and actual radial positions of tracers (redshift-space distortions; see \eg \citealt{Mao2014} for the impact of this on moments of the density field and \citealt{Uhlemann2018a} for the impact in the density PDF). Finally, both photometric and spectroscopic redshift measurements will only have a finite precision \citep[see][for recent work on photometric redshift estimation]{Hoyle2018}.

A way to circumvent these problems is to consider 2-dimensional projections of the density field instead - \eg the convergence of gravitational lensing which is a line-of-sight projection of the matter density field whose PDF can be modelled with approaches similar to the one developed here \citep{Bernardeau1995, Bernardeau2000, Barthelemy2019}. But even the PDF of tracer galaxies from photometric galaxy surveys can be successfully analysed with a formalism related to the one presented here \citep{Friedrich2018, Gruen2018}. To modify our 3D formalism accordingly, one has to consider cylindrical (as opposed to spherical) collapse in our derivations of \secrefnospace{model}. We defer this adjustment to future work.

\subsection{Baryonic feedback, tracer bias and stochasticity}

In this paper we have considered dark-matter-only simulations and neglected the impact of baryonic effects on the late-time evolution of structures \citep[see \eg][]{Schneider2019}. \citet{Foreman2019} find that baryonic physics leaves the Fourier space kernel that relates the late-time matter bispectrum and the late-time power spectrum mostly untouched, suggesting that baryonic feedback can be propagated into higher order correlators through its impact on the 2-point function. The latter can \eg be calibrated with hydrodynamic simulations, see \citet{Eifler2015, Huang2019} and references therein. For the modelling of the density PDF this would mean that baryonic physics can be largely incorporated through the variance re-scaling described in \secrefnospace{non_linear_variance_rescaling}. Alternatively, \citet{Ivanov2019} have investigated effective field theory corrections to their model of the matter density PDF. This may also be suitable to incorporate baryonic effects \cite[see \eg][who investigate this for the power spectrum]{Lewandowski2015}. This would again require calibration with simulations. The safest strategy to avoid the challenges of modelling baryonic feedback (and in fact any strongly non-linear evolution) is to analyse the PDF at high redshifts, \eg with Quasars as tracer samples.

When inferring the matter density PDF from a tracer sample (\eg galaxies) another major obstacle is the fact that there is in general only a biased and stochastic relationship between the tracer density contrast and the total matter density contrast (tracer bias and tracer stochasticity; see \citealt{Friedrich2018} or \citealt{Uhlemann2018a} for their impact of the galaxy density PDF and \citealt{Mao2014} for their impact on moments of the galaxy density field as well as \citealt{Dekel1999, Desjacques2018} for general introductions).

This problem can be circumvented entirely by directly studying the PDF of the convergence of gravitational lensing \citep{Barthelemy2019} since it directly measures the matter density field. An alternative ansatz pursued by \citet{Friedrich2018} and \citet{Gruen2018} is to jointly study counts-in-cells (\ie the galaxy density PDF) and lensing-around-cells. In early data of the Dark Energy Survey, they measured the parameters of a stochastic galaxy bias model and could still infer information about both variance and skewness of the matter density field.

\section*{Acknowledgements} 

OF gratefully acknowledges support by the Kavli Foundation and the International Newton Trust through a Newton-Kavli-Junior Fellowship and by Churchill College Cambridge through a postdoctoral By-Fellowship. CU kindly acknowledges funding by the STFC grant RG84196 `Revealing the Structure of the Universe'. Part of the work of FVN has been supported by the Simons Foundation. TB acknowledges support from the Cambridge Center for Theoretical Cosmology through a Stephen Hawking Advanced Fellowship. MM acknowledges support from the European Union's Horizon 2020 research and innovation program under Marie Sklodowska-Curie grant agreement No 6655919Y. TN was supported in part by JSPS KAKENHI Grant Numbers JP17K14273 and 19H00677, and by the Japanese Science and Technology Agency CREST JPMHCR1414. We would like to thank William Coulton, Daniel Gruen, Tim Eifler, Enrico Pajer, Blake Sherwin, Sandrine Codis, Omar Darwish, Paul Shellard, Anthony Challinor, George Efstathiou, Vid Irsic, Matteo Biagetti, Chihway Chang and Bhuvnesh Jain for helpful comments and discussions. We would also like to thank Kacper Kornet for HPC support as well the anonymous journal referee for helpful comments.

\section*{Data availability} 

\verb|C++| and \verb|python| tools to compute our model predictions are publicly available at \url{https://github.com/OliverFHD/CosMomentum} . Summary statistics measured in the Quijote N-body simulations are publicly available at \url{https://github.com/franciscovillaescusa/Quijote-simulations} .
\newpage

\def\aj{AJ}%
\def\araa{ARA\&A}%
\def\apj{ApJ}%
\def\apjl{ApJ}%
\def\apjs{ApJS}%
\def\ao{Appl.~Opt.}%
\def\apss{Ap\&SS}%
\def\aap{A\&A}%
\def\aapr{A\&A~Rev.}%
\def\aaps{A\&AS}%
\def\azh{AZh}%
\def\baas{BAAS}%
\def\jrasc{JRASC}%
\def\memras{MmRAS}%
\def\mnras{MNRAS}%
\def\pra{Phys.~Rev.~A}%
\def\prb{Phys.~Rev.~B}%
\def\prc{Phys.~Rev.~C}%
\def\prd{Phys.~Rev.~D}%
\def\pre{Phys.~Rev.~E}%
\def\prl{Phys.~Rev.~Lett.}%
\def\pasp{PASP}%
\def\pasj{PASJ}%
\def\qjras{QJRAS}%
\def\skytel{S\&T}%
\def\solphys{Sol.~Phys.}%
\def\sovast{Soviet~Ast.}%
\def\ssr{Space~Sci.~Rev.}%
\def\zap{ZAp}%
\def\nat{Nature}%
\def\iaucirc{IAU~Circ.}%
\def\aplett{Astrophys.~Lett.}%
\def\apspr{Astrophys.~Space~Phys.~Res.}%
\def\bain{Bull.~Astron.~Inst.~Netherlands}%
\def\fcp{Fund.~Cosmic~Phys.}%
\def\gca{Geochim.~Cosmochim.~Acta}%
\def\grl{Geophys.~Res.~Lett.}%
\def\jcap{JCAP}%
\def\jcp{J.~Chem.~Phys.}%
\def\jgr{J.~Geophys.~Res.}%
\def\jqsrt{J.~Quant.~Spec.~Radiat.~Transf.}%
\def\memsai{Mem.~Soc.~Astron.~Italiana}%
\def\nphysa{Nucl.~Phys.~A}%
\def\physrep{Phys.~Rep.}%
\def\physscr{Phys.~Scr}%
\def\planss{Planet.~Space~Sci.}%
\def\procspie{Proc.~SPIE}%

\bibliographystyle{mnras}
\bibliography{literature}

\appendix

\section{Spherical collapse in $\Lambda$CDM}
\label{app:spherical_collapse}

In the Newtonian approximation and setting $G = 1 = c$ the evolution of spherical, cylindrical or planar perturbations $\delta$ is described by
\begin{equation}
\label{eq:SC_in_conformal_time}
\ddot{\delta} + \mathcal{H} \dot{\delta} - \frac{N+1}{N} \frac{\dot{\delta}^2}{1+\delta} \ = 4\pi \bar \rho_m a^2 \delta (1+\delta)\ ,
\end{equation}
where $\tau$ is conformal time, $\mathcal{H} = \dd \ln a / \dd \tau$ is the conformal expansion rate and $N=3$ for a spherical perturbation, $N=2$ for a cylindrical perturlation and $N=1$ for a planar perturbation (see \citet{MukhanovBook} who demonstrates this for $N=1$ and $N=3$). To evaluate \eqnref{delta_R_equals_F} and related expressions we choose $N=3$ and solve \eqnref{SC_in_conformal_time} with the initial conditions
\begin{equation}
    \delta_i = \delta_{L,R_L}^*\ D(z_i)\ ,\ \dot{\delta_i} = \delta_i\ \mathcal{H}(z_i)\ ,
\end{equation}
where $z_i$ is a redshift chosen during matter domination. (In fact, in our calculation of $D(z)$ we set the radiation density $\Omega_r$ to zero and then choose $z_i = 4000$.)

\section{Fourier space conventions}
\label{app:Fourier_conventions}

We define the Fourier transform of a function $f(\mathbf{x})$ in real space as
\begin{equation}
    \tilde f(\mathbf{k}) := \int \dd^3 x\ f(\mathbf{x})\ e^{-i\mathbf{x}\mathbf{k}}\ . 
\end{equation}
In this convention, the convolution theorem takes the form
\begin{equation}
    \widetilde{f*g}(\mathbf{k}) = \tilde f(\mathbf{k})\tilde g(\mathbf{k})\ ,\ \widetilde{fg}(\mathbf{k}) = \frac{1}{(2\pi)^3}(\tilde f*\tilde g)(\mathbf{k})\ .
\end{equation}
And the power spectrum of the density contrast field $\tilde\delta(\mathbf{k})$ is given by
\begin{align}
    \langle \tilde\delta(\mathbf{k})\tilde\delta(\mathbf{q})\rangle = (2\pi)^3 \delta_D(\mathbf{k}+\mathbf{q})\ P(k)\ .
\end{align}
For the 2-point correlation function of the real space density contrast $\delta(\mathbf{x})$ this means that
\begin{align}
    \xi(\mathbf{r}) &\ =\langle \delta(\mathbf{x}+\mathbf{r}) \delta(\mathbf{x})\rangle\nonumber  = \int \frac{\dd^3k\ \dd^3q}{(2\pi)^6}\ \langle \tilde{\delta}(\mathbf{k})\tilde{\delta}(\mathbf{q})\rangle\ e^{i\mathbf{k}(\mathbf{x}+\mathbf{r})}e^{i\mathbf{q}\mathbf{x}}\nonumber \\
    &\ = \int \frac{\dd^3k}{(2\pi)^3}\ P(k)\ e^{i\mathbf{k}\mathbf{r}}\ ,
\end{align}
\ie it is the Fourier transform of the power spectrum. The variance of the density field when averaged over a top-hat filter of radius $R$ is given by
\begin{align}
    \langle \delta_{L,R}^2 \rangle =\ & \int \frac{\dd^3k_1\dd^3k_2}{(2\pi)^6}\ \langle \tilde \delta(\mathbf{k}_1) \tilde \delta(\mathbf{k}_2)\rangle\ \tilde W_R(k_1) \tilde W_R(k_2) \nonumber \\
    =\ & \frac{1}{2 \pi^2}\int \dd \ln k\ k^3\ P_L(k)\ \tilde W_R(k)^2 \ ,
\end{align}
where $\tilde W_R(k)$ is given in \eqnref{filter}.

\section{Testing the model accuracy with simulations of even higher resolution}
\label{app:higher_res_test}

In \figref{Quijote_residuals} we had complemented the N-body simulations described in \secref{sims} with a set of simulations run by \citet{Baldauf2016}. These dark-matter-only simulations consist of 14 boxes of volume $V=(1.5 \mathrm{Gpc}/h)^3$ sampled by $1024^3$ matter particles. The spacing between particles in the initial density field of these simulations is hence $\approx 1.5 \mathrm{Mpc}/h$, which places them between the resolutions of the fiducial Quijote runs and the high-resolution Quijote runs.

\begin{centering}
\begin{figure}
  \includegraphics[width=0.47\textwidth]{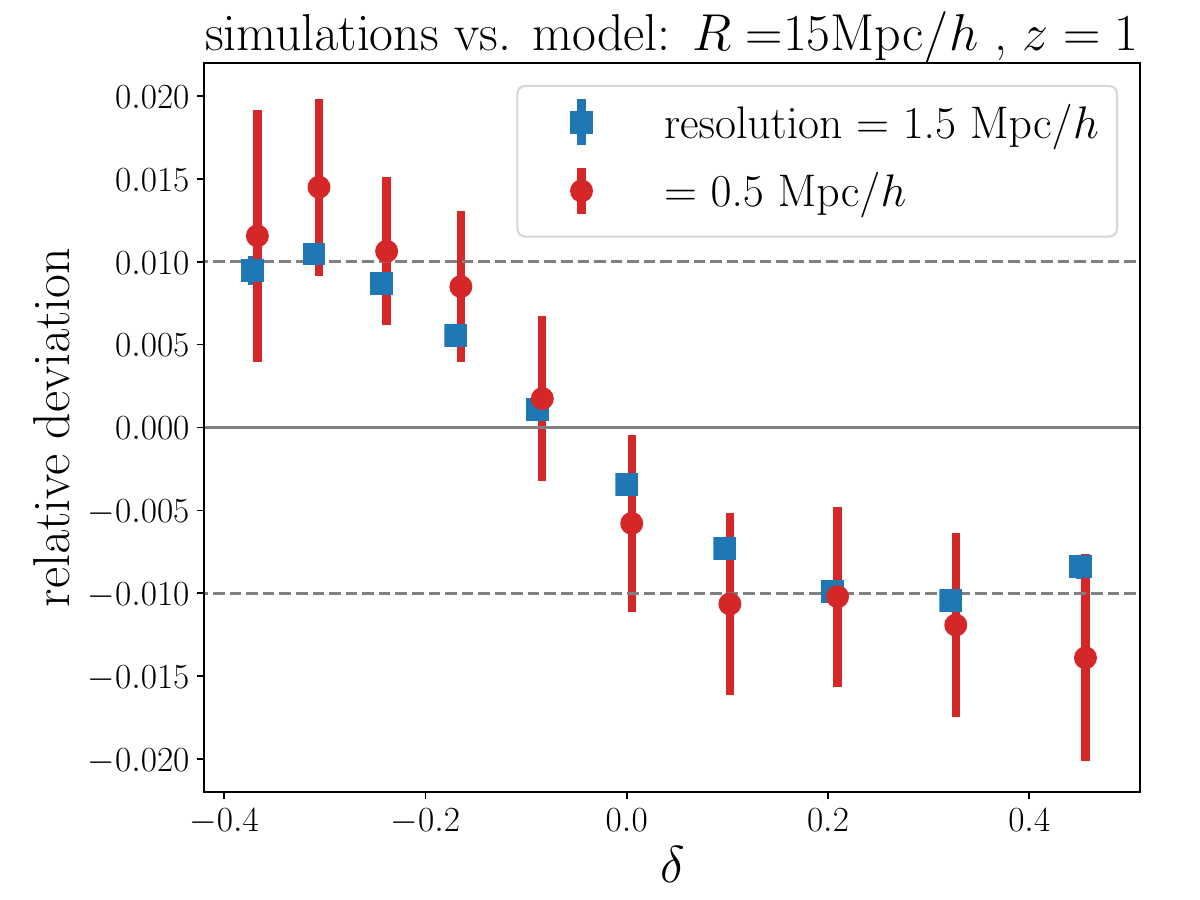}
   \caption{Same as the lower panel of \figref{Quijote_residuals} but using only the simulations run by \citet{Baldauf2016} and a set of simulations with the same cosmology, but resolution that is even higher than the high-resolution runs of Quijote. Going to this high resolution does not seem to improve agreement between model and simulations any further. Though the amount of high resolution simulations we were able to run is limited (the errorbars indeed denote uncertainty on the mean of our 6 runs) and we will investigate this further in the future.}
  \label{fi:Tobias_residuals}
\end{figure}
\end{centering}

To push our test of the impact of simulations resolution to even higher resolutions we ran another 5 N-body boxes at the same cosmology and with the same number of particles as in \citet{Baldauf2016} but with a volume of $V=(0.5 \mathrm{Gpc}/h)^3$ instead. In \figref{Tobias_residuals} we show the relative deviation between the PDF measured at $R=15\mathrm{Mpc}/h$ and $z=1$ and our analytic prediction - again computed for the best-fit non-linear variance. We also show the lower resolution version of these simulations in the figure. It seems that going to resolutions that are even higher than that of the Quijote high-resolution runs (blue points in \figrefnospace{Quijote_residuals}) does not further improve the agreement between our model and the PDF of the simulated density field. However, the statistical uncertainty in the PDF measurements obtained from these 5 small boxes is substantial and we are going to investigate this further with larger sets of high-resolution simulations in the future.
\section{Simultaneously varying different Bispectrum shapes}
\label{app:vary_different_shapes}
\begin{figure*}
\begin{center}
  \includegraphics[width=0.9\textwidth]{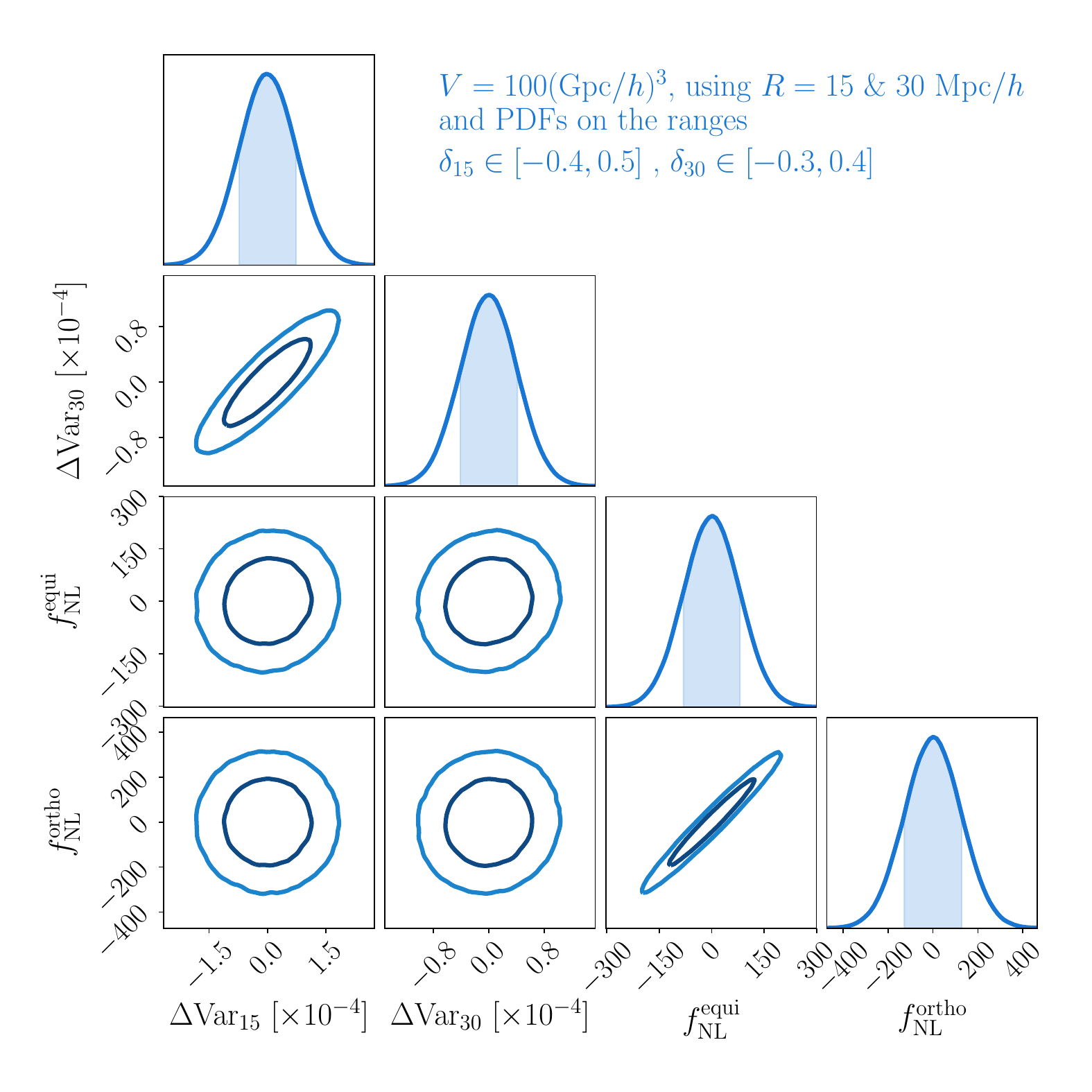}
   \end{center}
   \caption{Same as figure \figref{constraints_joint} but simultaneously varying $f_{\mathrm{NL}}^{\mathrm{equi}}$ and $f_{\mathrm{NL}}^{\mathrm{ortho}}$. The strong degeneracy between these two parameters for both the moment and PDF analysis demonstrates that circular apertures are not well suited to distinguish between different primordial Bispectrum shapes.}
  \label{fi:constraints_joint_equi_ortho}
\end{figure*}

In general, the primordial bispectrum could be a super position of different templates. \Eg \citetalias{Planck2018PNG} simultaneously varied $f_{\mathrm{NL}}^{\mathrm{equi}}$ and $f_{\mathrm{NL}}^{\mathrm{ortho}}$ in their analysis. In \figref{constraints_joint_equi_ortho} we forecast the same measurements as displayed in \figref{constraints_joint} but this time simultaneously vary $f_{\mathrm{NL}}^{\mathrm{equi}}$ and $f_{\mathrm{NL}}^{\mathrm{ortho}}$. The strong degeneracy between these two parameters in a PDF analysis demonstrates that circular apertures are not well suited to distinguish between different primordial Bispectrum shapes. However, note that the analysis of \citetalias{Planck2018PNG} finds very little degeneracy between orthogonal and equilateral bispectrum shapes. Hence an analysis of late-time density PDFs would still provide complementary information, even when trying to measure $f_{\mathrm{NL}}^{\mathrm{equi}}$ and $f_{\mathrm{NL}}^{\mathrm{ortho}}$ simultaneously. Also, an analysis of the PDF at a wider range of scales would be more sensitive to the scale dependence of the primordial skewness and hence help to further disentangle between different bispectrum shapes.

\section{Assumption of Gaussian likelihood of PDF measurements}
\label{app:Gaussian_likelihood}

Throughout this paper we employed the assumption that statistical uncertainties in PDF measurements follow a multivariate Gaussian distribution. This cannot be completely true for two reasons: First, PDF measurements will always be positive which necessarily skews their distribution. This is especially noticeable in the tails of the PDF, where sampling noise is expected to lead to a Poisson-like rather than a Gaussian distribution. Secondly, histograms of density fluctuations that measure the density PDF are normalised such that integration over these histograms gives $1$. This means that the bins in which the PDF has been measured cannot be perfectly independent of each other, which seems to further exclude a multivariate Gaussian distribution.

\begin{figure*}
\begin{center}
  \includegraphics[width=0.49\textwidth]{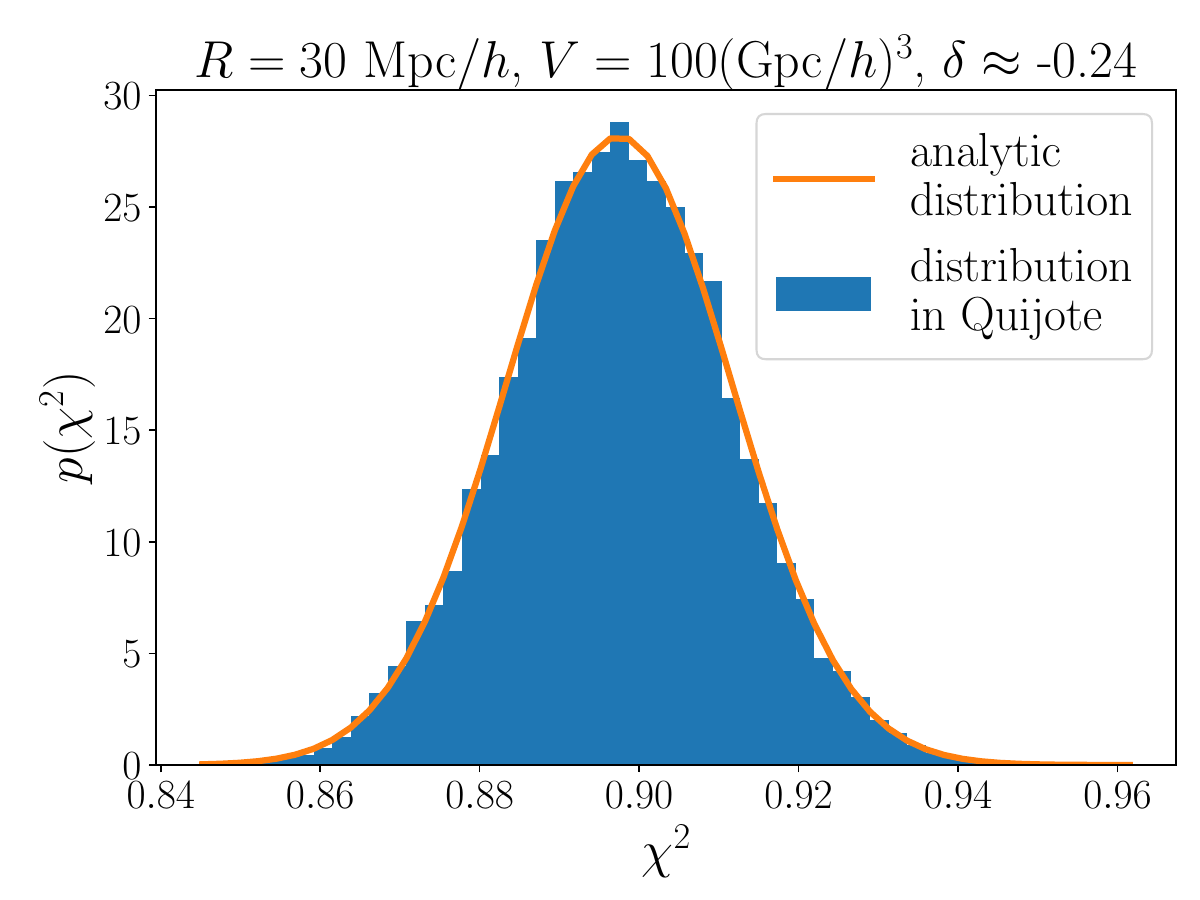}\includegraphics[width=0.49\textwidth]{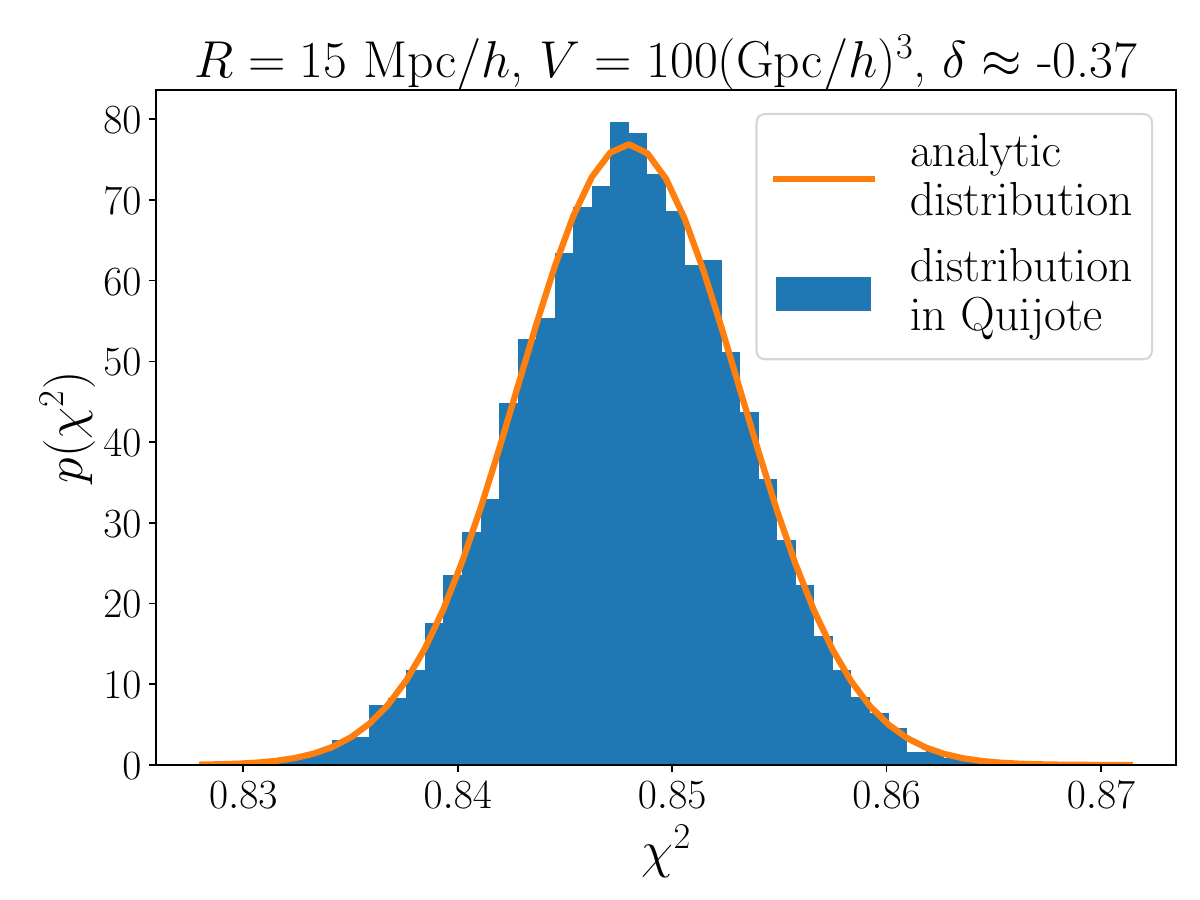}
   \end{center}
   \caption{The blue histograms show the distribution of individual PDF measurements in the Quijote simulations at smoothing scales of $R=30$Mpc$/h$ (left panel) and $R=15$Mpc$/h$ (right panel) and at the lowest densities that went into our analyses. These density bins are the ones that show the strongest deviation from a Gaussian distribution. The orange lines display analytic Gaussian distributions with the same mean and variance.}
  \label{fi:data_histograms}
\end{figure*}

However, in our analysis have cut the tails of the density PDF which reduces both of these problems. We demonstrate this in \figref{data_histograms} where we show histograms of PDF measurements in the Quijote simulations for the lowest density bins that went into our analysis of either the $R=30$Mpc$/h$ and $R=15$Mpc$/h$ smoothing scale. The figure also shows analytic Gaussian distributions with the same mean and variance. There is a close agreement of these analytic distributions with the actual distributions in Quijote. This indicates that even for the bins in our analysis that reach farthest into the tails, the measurement uncertainties are close to Gaussian. We have checked that this also holds for all other density bins in our analysis.

\begin{figure*}
\begin{center}
  \includegraphics[width=0.49\textwidth]{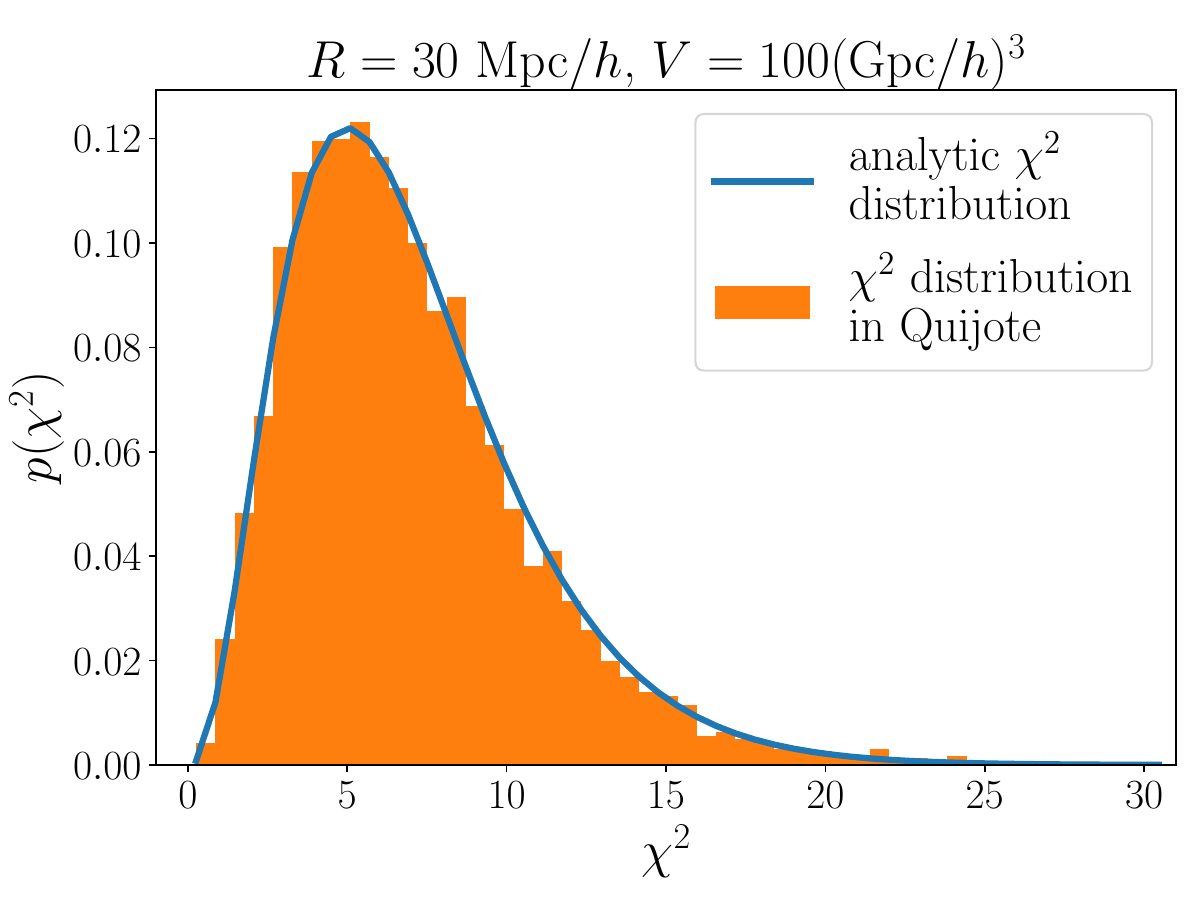}\includegraphics[width=0.49\textwidth]{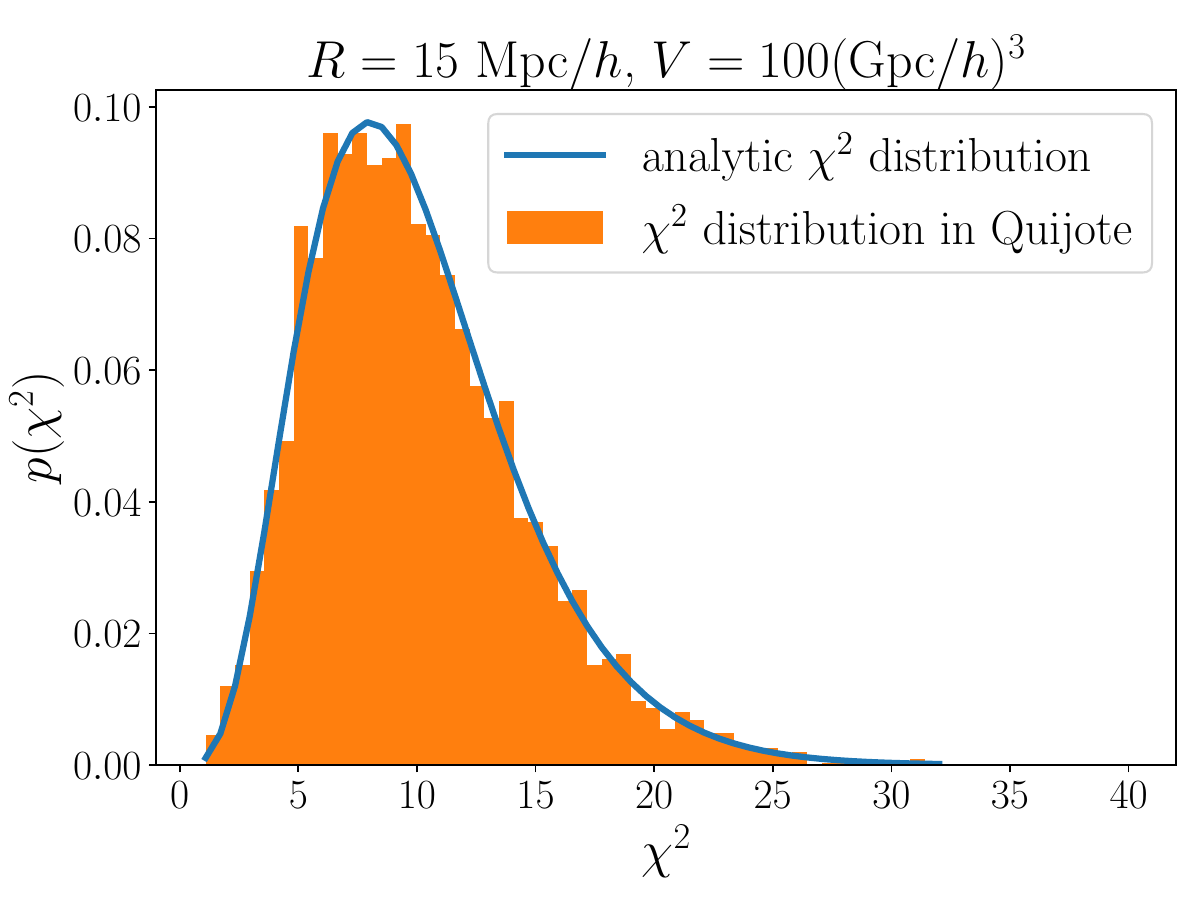}
   \end{center}
   \caption{The orange histograms show the distribution of $\chi^2$ between the mean PDF measured in $5000$ Quijote simulations and the individual PDF measurements in these simulations at smoothing scales of $R=30$Mpc$/h$ (left panel) and $R=15$Mpc$/h$ (right panel). The blue lines display the analytic $\chi^2$ distributions that would be expected if statistical uncertainties in the PDF measurements were drawn from a multi-variate Gaussian distribution.}
  \label{fi:chiSq_histograms}
\end{figure*}

To test for the assumption of multivariate Gaussianity we also look at the $\chi^2$ of our entire PDF measurements. To do so we use $10.000$ of the Quijote simulations to estimate the covariance of the PDF measurments and then use that covariance to compute the $\chi^2$ of the remaining $5.000$ Quijote PDF measurements \wrt their mean measurement. The histograms of these $5.000$ $\chi^2$ values is shown in \figref{chiSq_histograms} and again there is a good agreement with the analytic expectation of a (multivariate) Gaussian distribution. Note that with $10.000$ simulations any noise in the covariance estimate (and its inverse) will be negligible in this test \citep{Taylor2013, Friedrich_Eifler}.

\end{document}